\documentclass[journal]{IEEEtran}
\usepackage{amsmath}

\usepackage[]{graphicx}% remove the 'demo' option in the real document

\usepackage{hyperref}% add hypertext capabilities

\ifCLASSINFOpdf
\else
\fi
\ifCLASSOPTIONcompsoc
  \usepackage[caption=false,font=normalsize,labelfont=sf,textfont=sf]{subfig}
\else
  \usepackage[caption=false,font=footnotesize]{subfig}
\fi
\begin{document}
\title{Wide-scanning Circularly Polarized Reflector-based Modulated Metasurface Antenna Enabled by a Broadband Polarizer}
%Wide-band/Wide-scanning Circularly Polarized Modulated Metasurface Antenna (MoMetA) Enabled by Angular Stable Polarizer
%High Gain and Wide-scanning Reflector Based Modulated Metasurface  Antenna with Circular Polarization
%Embedding Multi-layer Polarization Converter on Reflector Based Modulated Metasurface Antenna to Achieve High Gain and Wide Scanning Circularly Polarized Radiation

\author{Ali Mohammad Hakimi,
        Ali Keivaan,
        Homayoon Oraizi,~\IEEEmembership{Life~Senior Member,~IEEE}, and Amrollah Amini% <-this % stops a space
\thanks{A. M. Hakimi, A. Keivaan, H. Oraizi and A. Amini are with the School
of Electrical Engineering, Iran University of Science and Technology, Tehran 1684613114, Iran, (email: h\_ oraizi@iust.ac.ir)}
}

\maketitle

\begin{abstract}
A circularly polarized leaky-wave antenna capable of frequency scanning is proposed in this paper. The main objective is to achieve high gain and polarization purity without the need for a complex feed network. The antenna consists of two independent modules: (1) Anisotropic modulated metasurface antenna (MoMetA) for generating vertically polarized radiation with an operational scanning range of 19 to 47 degrees in elevation; (2) Wide-band polarization converter with high angular stability and capable of converting vertically linear polarization into circular polarization (CP). 
Aperture field estimation (AFE) method is used to design the antenna with high accuracy in predicting far-field pattern without the need for full-wave simulations. The gain of antenna in the bandwidth of 16 to 21 GHz is obtained better than 18 dBi. Simulation results show that the axial ratio in the maximum gain direction is lower than 3 dB all over its operational frequency bandwidth, demonstrating the proper operation of polarizer. In order to verify the simulation results, one prototype of the antenna is fabricated and its radiation patterns are measured in an anechoic chamber.
\end{abstract}
\IEEEpeerreviewmaketitle
\section{Introduction}
\IEEEPARstart{P}{lanar} leaky-wave antennas are suitable choices for radiators of radars and future generations of telecommunication systems. One merit of leaky-wave structures is their ability of frequency scanning in wide-band frequency ranges, which eliminates the need for phase array antennas with complex feed networks. Moreover, in satellite communication systems and frequency modulated continuous wave (FMCW) radars it is preferable to use circular polarization for data reception and transmission to avoid multipath interferences \cite{shafai2015}.
In recent years, different structures for realizing leaky-wave antennas with circular polarization and continuous frequency scanning have been proposed. Planar slot array antennas based on substrate integrated waveguides (SIW) \cite{jin2013}-\cite{chen2019}, periodic printed dipole arrays \cite{sanchez2016}, periodic patches \cite{fu2012}, \cite{rahmani2017}, spoof surface plasmon polariton (SSPP) based structures \cite{zhang2018} and composite right/left-handed (CRLH) antennas \cite{fu2017} have been proposed to realize such radiators. The mentioned references generally do not possess a high gain pencil beam radiation pattern, which is rooted in their structural geometry. In order to achieve a pencil beam radiation by these structures, an array in their transverse layouts is required. However, in such cases the complexity and the limited bandwidth of feed networks are the two obstacles imposed on their implementation.
 Alternatively, the modulated metasurface antennas (MoMetAs) make it possible to achieve high gain and desired polarization without the need for complex feed networks. In \cite{fong2010}, holographic technique was proposed to design vertically and circularly polarized leaky-wave antennas, which can be realized by printing patterned sub-wavelength patches on a dielectric slab. Consequently, the printed circuit board technology as a low cost and efficient option makes them suitable for integrated systems. The use of holographic technique helps to predict the wave-front direction radiated from the modulated metasurface. This method is also proposed for the synthesis of multi-beam leaky-wave antennas \cite{li2014}, \cite{movahhedi2019}, dual-functional antennas \cite{li2015}, dual-polarized antennas \cite{movahhedi2020} and electrically reconfigurable leaky-wave structures \cite{cheng2016}.
Other ways to design MoMetAs are the aperture field estimation (AFE) method \cite{teniou2017}, \cite{minatti2016} and adiabatic Floquet-wave analysis \cite{minatti_FO2016}.
The advantage of these methods over the holographic technique is their more accurate realization of radiation patterns and polarization type.
Although MoMetAs have the mentioned merits, but they suffer from narrow bandwidths and narrow frequency scanning range. In \cite{moeini_PRA2019} and \cite{moeini2019}, surface wave reflectors are utilized to incorporate scannability to the isotropic holograms. Surface wave reflectors can eliminate the destructive effects of backward modes. In other words, the less backward mode excitation results in a wider operation frequency range \cite{nanneti2007}.

In this paper, a leaky-wave structure with circular polarization and frequency scanning capability is designed  based on modulated metasurfaces. The antenna consists of two main parts: (1) Anisotropic MoMetA with vertically polarized radiation and frequency scannability; (2) Polarization converter which is located at the top of metasurface and converts vertical polarization into circular one. The AFE method is used to design the MoMetA. 
The common anisotropic MoMetAs do not possess scannability, and a modification in their structure is required to address this problem.  For this purpose, a wedge-shaped surface wave reflector with opening angle of 270 degrees on the hologram is implemented to hinder the excitation of non-forward leaky modes and to avoid their destructive effects. In order for the antenna to have an acceptable operation, the following specifications should be addressed:
(1) The antenna should have a high gain and the surface wave should be ideally converted into the space wave; (2) Cross-polarization level should be low, as much as possible, to avoid unwanted reflections from the polarizer surface; (3) The gain of antenna and cross-polarization level should not change as much as possible by varying the frequency to achieve the radiation scannability with circular polarization.
In the design of polarizer, its characteristics are obtained under ideal plane wave illumination with vertical polarization. Three main points are considered in the design of polarizer layer: (1) The polarizer should have a high power efficiency over the operational bandwidth, that is the insertion loss should be minimum; (2) The polarizer should have a high polarization purity over the operational bandwidth, that is the cross-polarization should be minimum; (3) The polarizer should retain its stability over the incident angle, since the direction of radiating beam is scanned by the frequency variation. 
\section{Antenna design}
\subsection{Aperture field estimation method}
\begin{figure}
\centering
\includegraphics[width=0.5\textwidth]{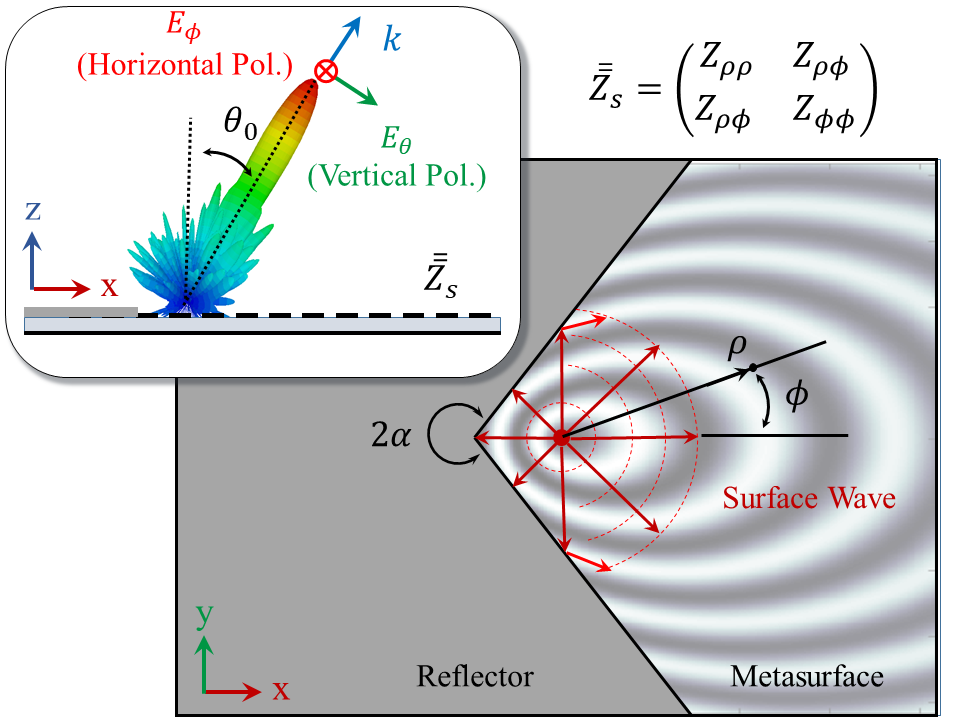}
\caption{The reflector-based anisotropic MoMetA consisting of monopole feed located at ($\rho_0, \phi_0$) in the cylindrical coordinates. The opening angle of the reflector is $2\alpha$.}
\label{fig:Fig1}
\end{figure}

\begin{figure}
%%%%
\centering
     \subfloat[\label{subfig-1:dummy}]{%
       \includegraphics[width=0.33\textwidth]{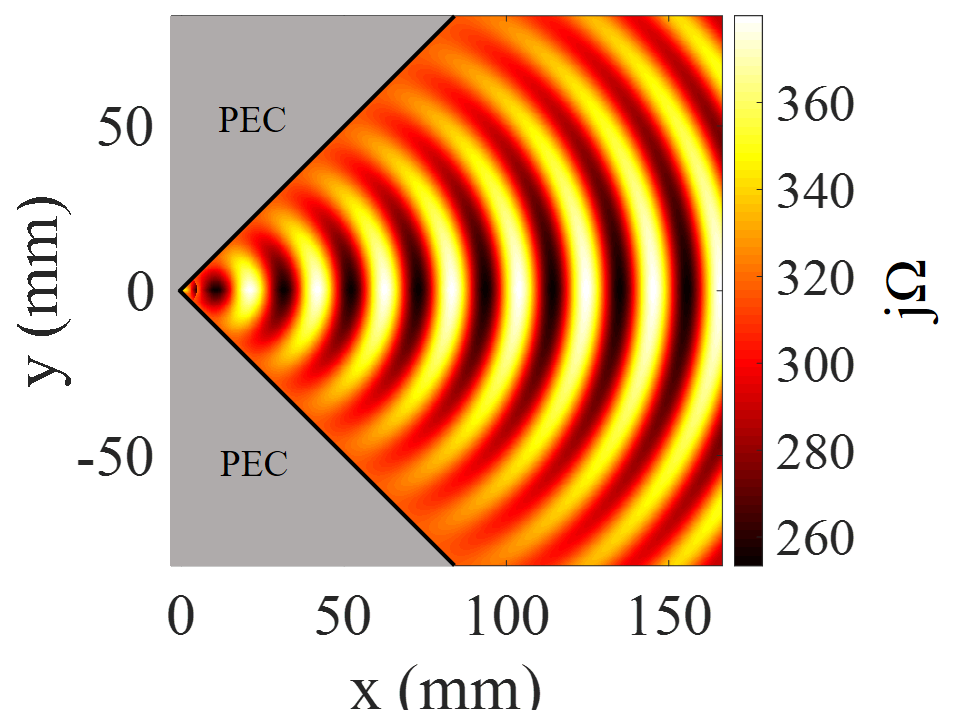}}\\
     %%%%%  
      \subfloat[\label{subfig-2:dummy}]{%
       \includegraphics[width=0.25\textwidth]{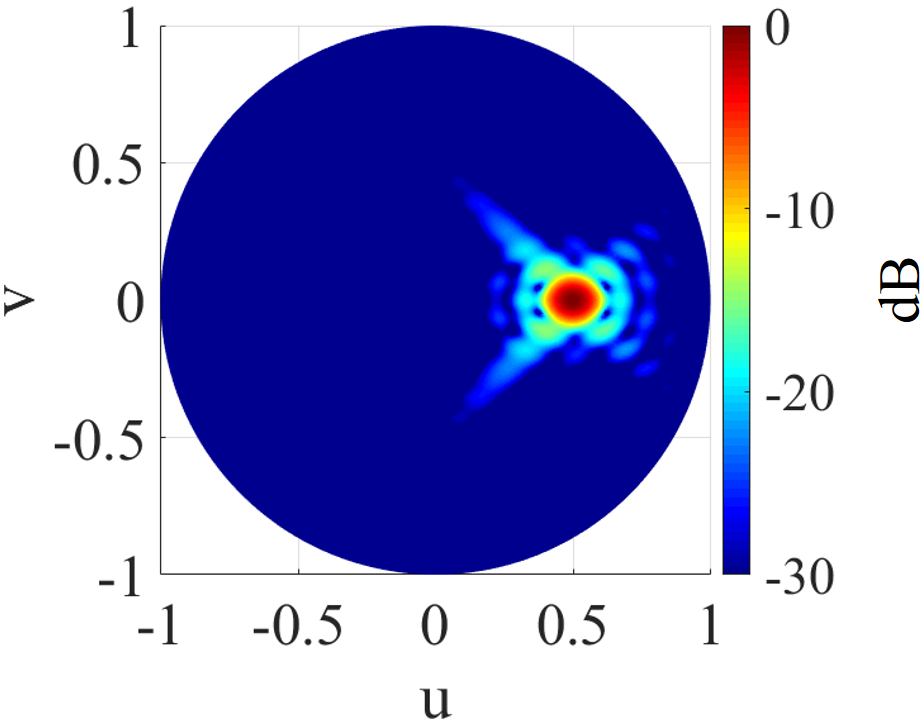}}
       %%%%%
     \subfloat[\label{subfig-3:dummy}]{%
       \includegraphics[width=0.25\textwidth]{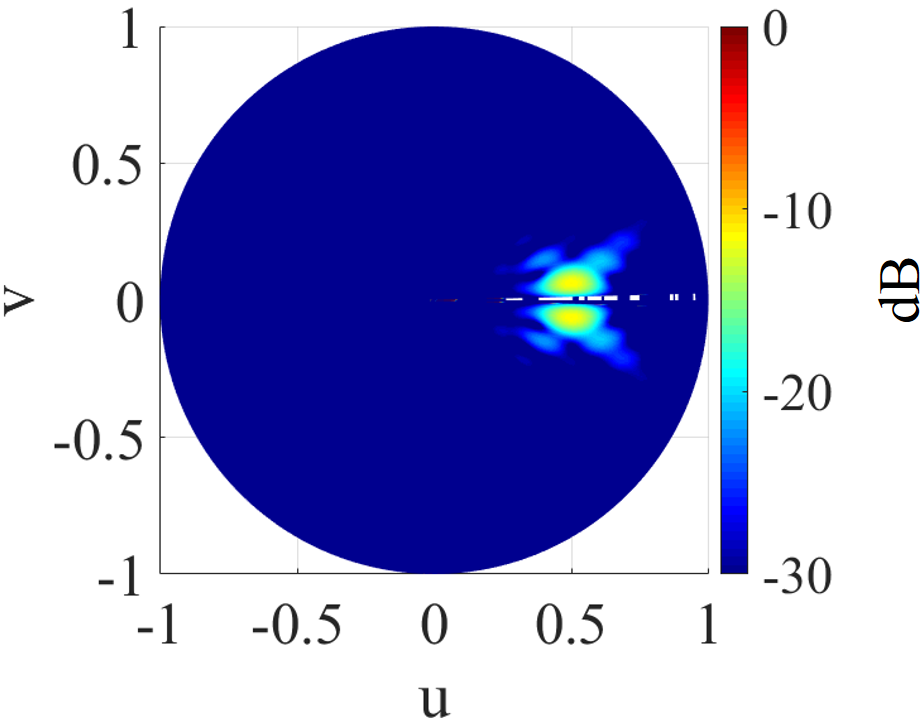}}
       %%%%%
       \caption{Surface reactance and radiation patterns of a reflector-based isotropic MoMetA (The opening angle $2\alpha$ is selected as 270 degrees). (a) $X_{\rho\rho}$, (b) vertical component ($F_\theta$) and (c) horizontal component ($F_\phi$).   }
\label{fig:Fig2}
\end{figure}

\begin{figure}
%%%%
     \subfloat[\label{fig:Fig3a}]{%
       \includegraphics[width=0.25\textwidth]{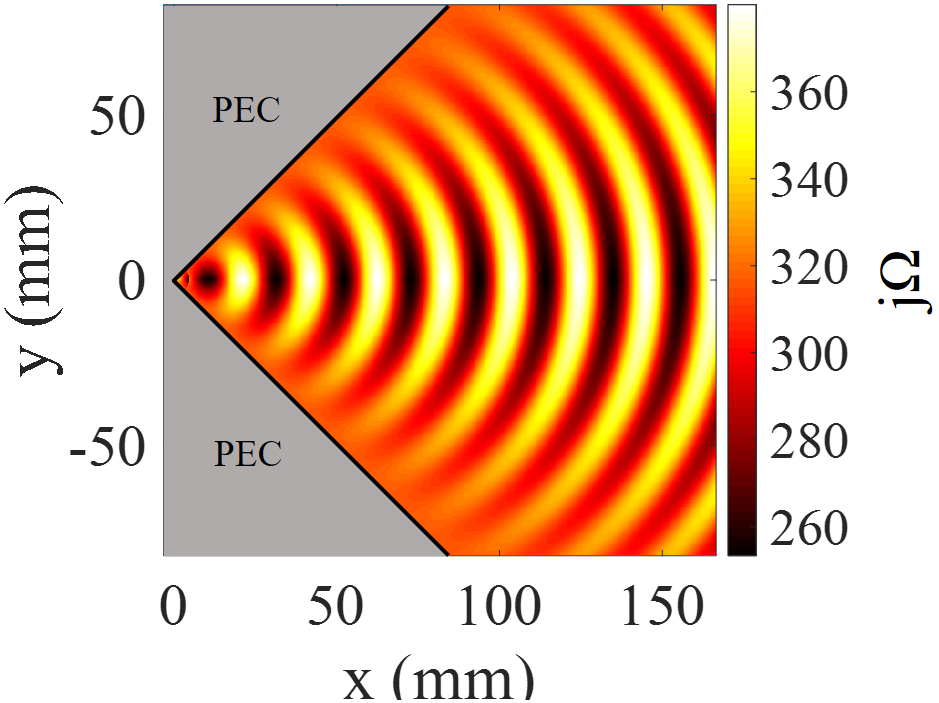}}
     %%%%%  
      \subfloat[\label{fig:Fig3b}]{%
       \includegraphics[width=0.25\textwidth]{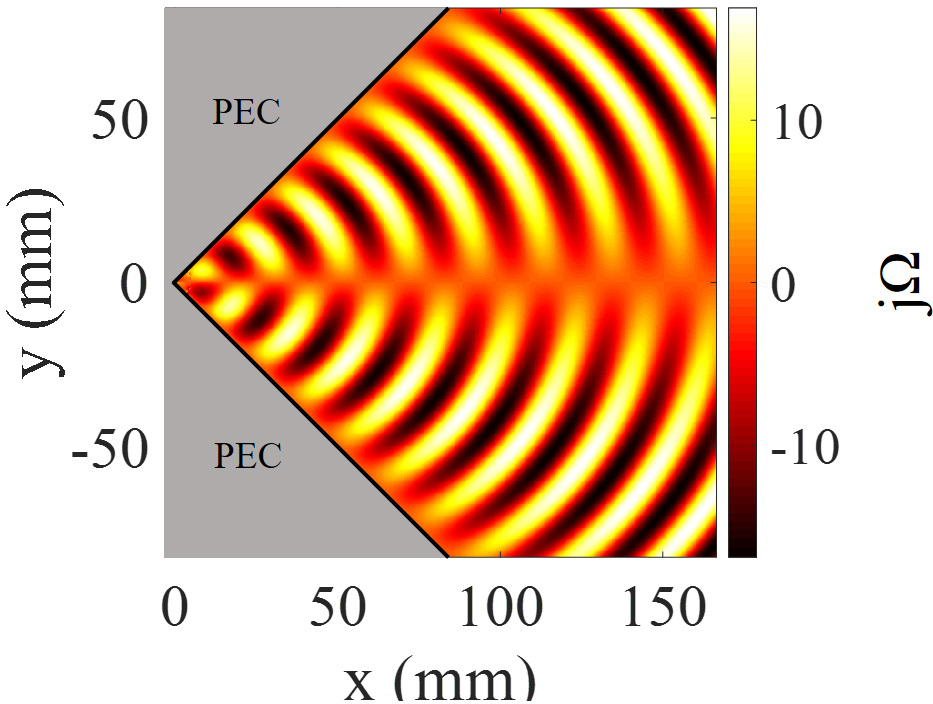}}\\
       %%%%%
     \subfloat[\label{fig:Fig3c}]{%
       \includegraphics[width=0.25\textwidth]{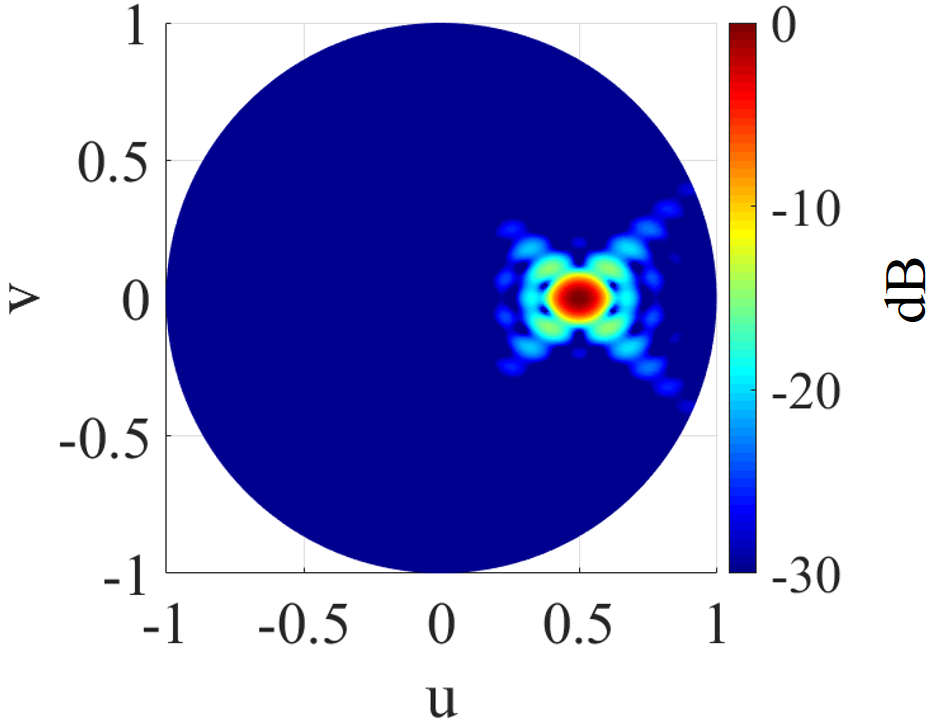}}
       %%%%%
       \subfloat[\label{fig:Fig3d}]{%
       \includegraphics[width=0.25\textwidth]{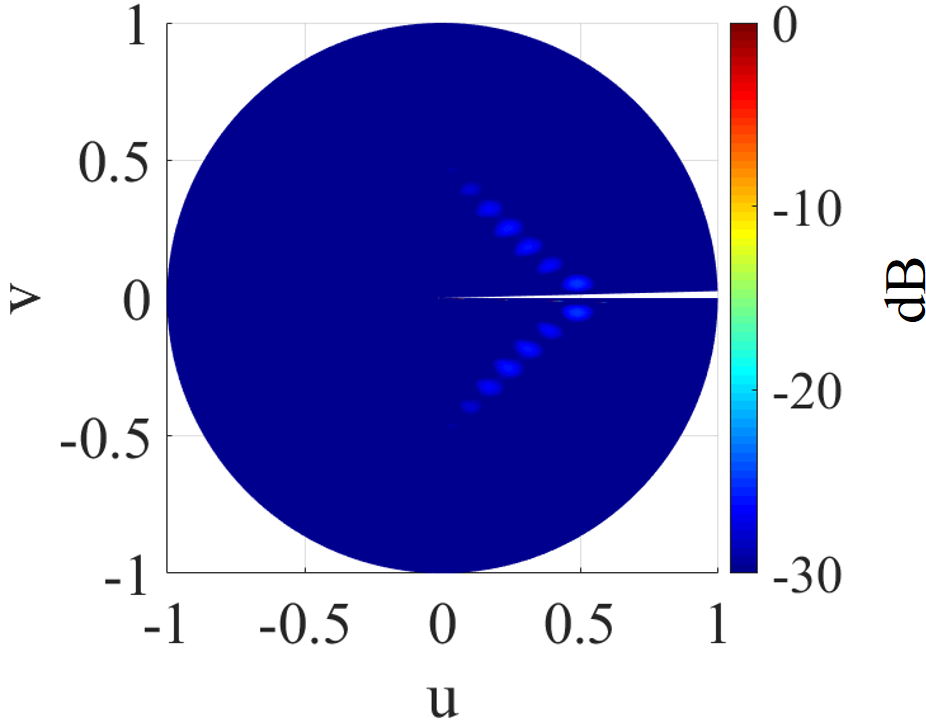}}
       \caption{Surface reactance and radiation patterns of a reflector-based anisotropic MoMetA (The opening angle $2\alpha$ are selected as 270 degrees). (a) $X_{\rho\rho}$, (b) $X_{\rho\phi}$, (c) vertical component ($F_\theta$) and (d) horizontal component ($F_\phi$).}
       \label{fig:Fig3}
\end{figure}

A classic method for the synthesis of planar leaky-wave antennas with linear polarization is employing the holographic technique, which was first practically proposed by Sievenpiper et al. \cite{sievenpiper2005}.
 In this technique, pseudo-periodic patches printed on dielectric substrates operate as an interferogram which are synthesized by the interference of a surface wave (or reference wave) as $\psi_{ref}$ and the desired wave-front (or object wave) as $\psi_{obj}$. The surface impedance required for the synthesis of object wave is calculated as \cite{fong2010}:
\begin{equation}
Z_s(\rho, \phi) = jX_0[1 + M\times Re(\psi_{ref}\psi_{obj}^*)]
\label{eq:Zs}
\end{equation}
where $X_0$ is the average surface reactance and $M$ is the modulation depth.  Equation (\ref{eq:Zs}) is used for the synthesis of isotropic impedance surfaces. The problem with the isotropic impedance surface is that the synthesized wave  has an acceptable polarization purity only in the design angle, but at the other angles of visible region, the cross-polarization is improper \cite{amini2020}.
Aperture field estimation method \cite{teniou2017} , \cite{minatti2016} is proposed for the synthesis of modulated impedance surfaces (as an hologram) to  control the desired polarization. The general relationship between the surface impedance tensor and aperture field $\vec{E}_a$ is defined as \cite{minatti2016}:
\begin{equation}
\bar{\bar{Z}}_s(\rho, \phi).\hat{\rho} = j[X_0\hat{\rho} + 2 Im(\frac{\vec{E}_a(\rho, \phi)}{\hat{\phi}.\vec{H}_t|_{z=0^+}(\rho, \phi)})] 
\label{eq:zs_tensor}
\end{equation}
where $\bar{\bar{Z}}_s(\rho, \phi)$ is the surface impedance tensor and is generally described by:
\begin{equation}
\bar{\bar{Z}}_s = Z_{\rho\rho}\hat{\rho}\hat{\rho} + Z_{\rho\phi}(\hat{\rho}\hat{\phi} + \hat{\phi}\hat{\rho}) + Z_{\phi\phi}\hat{\phi}\hat{\phi}
\end{equation}
Also, $\vec{H}_t$ is the magnetic field generated by the surface wave source. For the center-fed MoMetAs, the magnetic field is in the form of Hankel function of the second kind and first order \cite{minatti2016}:
\begin{equation}
\vec{H}_t|_{z=0^+} = -\hat{\phi}J_{sw}H_1^{(2)}(k_{sw}\rho)
\end{equation}
Note that $k_{sw}=\beta_{sw} - j\alpha_{sw}$ indicates the surface wave propagation constant. In all previous references, this form of feed is used to excite anisotropic holograms. The most common method to excite anisotropic holograms is to locate the feed at the center of antenna, yet it does not guarantee an acceptable radiation pattern across the specified bandwidth and it provides a proper pattern only at the design frequency. As a consequent, by getting away from the design frequency,  the pattern will be deteriorated caused by the destructive effects of backward modes \cite{moeini2019}, \cite{nanneti2007}. As a result, the backward modes will greatly reduce the operational bandwidth of antenna.
One suitable method for eliminating the adverse effects of backward modes and improving the bandwidth of the antenna is through implementing surface wave reflectors. In \cite{moeini_PRA2019}, \cite{moeini2019}, surface wave reflectors are used to increase the bandwidth and to realize the scannability in isotropic holographic antennas.
In this paper, a wedge-shaped reflector with opening angle of 270 degrees is used to enhance the bandwidth of anisotropic MoMetA. Hence, in addition to high polarization purity, the frequency scannability is also incorporated in the structure. Fig.\ref{fig:Fig1} shows the conceptual structure of reflector and the modulated metasurface.
Note that in the presence of the wedge-shaped reflector with opening angle of $2\alpha$, the form of magnetic field ($\vec{H}_t|_{z=0^+}=H_\phi$) should be modified as \cite{moeini2019}:
\begin{equation}
\vec{H}_\phi = -
\begin{cases}
\sum_\nu a_\nu J_\nu(k_{sw}\rho) H_\nu^{(2)}(k_{sw}\rho_0) S(\phi, \phi_0), \quad \rho < \rho_0\\
\sum_\nu a_\nu J_\nu(k_{sw}\rho_0) H_\nu^{(2)}(k_{sw}\rho) S(\phi, \phi_0), \quad \rho \geq \rho_0

\end{cases}
\label{eq:HH_phi}
\end{equation}
where $S(\phi, \phi_0)=\sin (\nu(\phi_0-\alpha)) \sin (\nu(\phi - \alpha))$ and $\nu$ and $a_\nu$ are defined as:
\begin{equation}
\nu = \frac{m\pi}{2(\pi - \alpha)} \quad m = 0, 1, 2, ...
\end{equation}
\begin{equation}
a_\nu = -\frac{\pi\omega \mu J_{sw}}{2(\pi - \alpha)}
\end{equation}
Also, $(\rho_0, \phi_0)$ indicates the place of feed in cylindrical coordinates. In (\ref{eq:zs_tensor}), the aperture field can be chosen arbitrarily, with the general form of:
\begin{equation}
\vec{E}_a(\rho, \phi) = E_{ax}(\rho, \phi)\hat{x} + E_{ay}(\rho, \phi)\hat{y}
\end{equation}
Thus, by properly choosing the x and y components of aperture field, the desired polarization in the far-field can be attained. Since the far-field  and aperture field are related by the Fourier transformation, the following equations are valid \cite{balanis2005}:
\begin{equation}
\vec{E}_{FF}(r, \theta, \phi) \approx \frac{jke^{-jkr}}{2\pi r} [F_\theta (\theta, \phi) \hat{\theta} + F_\phi (\theta, \phi) \hat{\phi}]
\end{equation}
\begin{equation}
F_\theta (\theta, \phi) = f_x \cos \phi + f_y \sin \phi
\end{equation}
\begin{equation}
F_\phi (\theta, \phi) = \cos \theta (-f_x \sin \phi + f_y \cos \phi)
\label{eq:F_phi}
\end{equation}
\begin{equation}
f_x(k_x, k_y) = \iint_{ap} E_{ax}(\rho ', \phi ')e^{j(k_x x' + k_y y')}\rho'd\rho' d\phi'
\label{eq:fx}
\end{equation}
\begin{equation}
f_y(k_x, k_y) = \iint_{ap} E_{ay}(\rho ', \phi ')e^{j(k_x x' + k_y y')}\rho'd\rho' d\phi'
\label{eq:fy}
\end{equation}
where $x' = \rho' \cos \phi'$ and $y' = \rho' \sin \phi'$. The integrations in (\ref{eq:fx}) and (\ref{eq:fy}) are performed on the effective aperture of metasurface.
In this paper, without loss of generality, the goal is to design a MoMetA which satisfies the following conditions:
\begin{enumerate}
\item The antenna should have a linearly polarized radiation pattern with high polarization purity at the design angle.
\item The beam angle at the desired frequency should be at $\theta = \theta_0$ and $\phi = \phi_0$.
\item The antenna should have frequency scannability over the specified bandwidth.
\end{enumerate}
The first condition determines the relation between x and y components of aperture field. For example, for having vertical polarization at the design angle, we should have:
\begin{equation}
F_\phi(\theta_0, \phi_0) = 0
\label{eq:FF_phi}
\end{equation}
Therefore, the relation between x and y components of Fourier transforms will be obtained by (\ref{eq:F_phi}) and (\ref{eq:FF_phi}):
\begin{equation}
f_{yy} = f_{xx} \tan \phi_0
\end{equation}
By the linearity of Fourier transformation, we have:
\begin{equation}
E_{ay} = E_{ax} \tan \phi_0
\end{equation}
However, for the beam to be in the direction of $(\theta_0, \phi_0)$, the x component of electric field should be defined as:
\begin{equation}
E_{ax}(\rho, \phi) = E_0(\rho, \phi) e^{-jk(x \sin \theta_0 \cos \phi_0 + y \sin \theta_0 \sin \phi_0)}
\end{equation}
The profile of $E_0(\rho, \phi)$ can be selected as desired \cite{minatti2016}. Given that the asymptotic form of Hankel functions in (\ref{eq:HH_phi}) is a function of $e^{-\alpha_{sw}\rho}$, it is convenient to define the profile of the field as a function of $e^{-\alpha_{sw}\rho}$, to keep the impedance modulation coefficient constant. Hence, the profile of electric filed can be selected as:
\begin{equation}
E_0(\rho, \phi) = M e^{-\alpha_{sw}\rho}
\end{equation}
where $M$ determines the depth of impedance modulation.
For the horizontal polarization, the equations are simplified as:
\begin{equation}
F_\theta(\theta_0, \phi_0) = 0
\end{equation}
\begin{equation}
E_{ax} = -E_{ay} \tan \phi_0
\end{equation}
To investigate the performance of anisotropic MoMetA, its far-field pattern is compared with the isotropic one. Note that for isotropic MoMetAs the aperture field may be defined as  \cite{amini2020}:
\begin{equation}
\vec{E}_a = E_0(\rho, \phi) e^{-jk(x \sin \theta_0 \cos \phi_0 + y \sin \theta_0 \sin \phi_0)} \hat{\rho}
\end{equation}
 Figs \ref{fig:Fig2} and \ref{fig:Fig3} show the synthesized surface reactances and calculated far-field patterns of the isotropic and anisotropic holograms, respectively. They are designed to radiate vertically polarized waves at 18 GHz and at $\theta_0 = 30^\circ$ and $\phi_0 = 0^\circ$. The parameters $X_0$ and $M$ are selected as $0.84\eta_0$ and $0.2$, respectively. Their values are chosen in a way that the required impedance range may be realized by our available technology.
 Note that, the far-field patterns are normalized to the maximum of co-polarized component. 
 Results show that, for anisotropic MoMetA the cross-polarized component is significantly reduced compared to the isotropic one all over the visible region.
\subsection{Synthesis of hologram with rectangular patches}
\begin{figure}
%%%%
\centering
     \subfloat[\label{fig:Fig4a}]{%
       \includegraphics[width=0.18\textwidth]{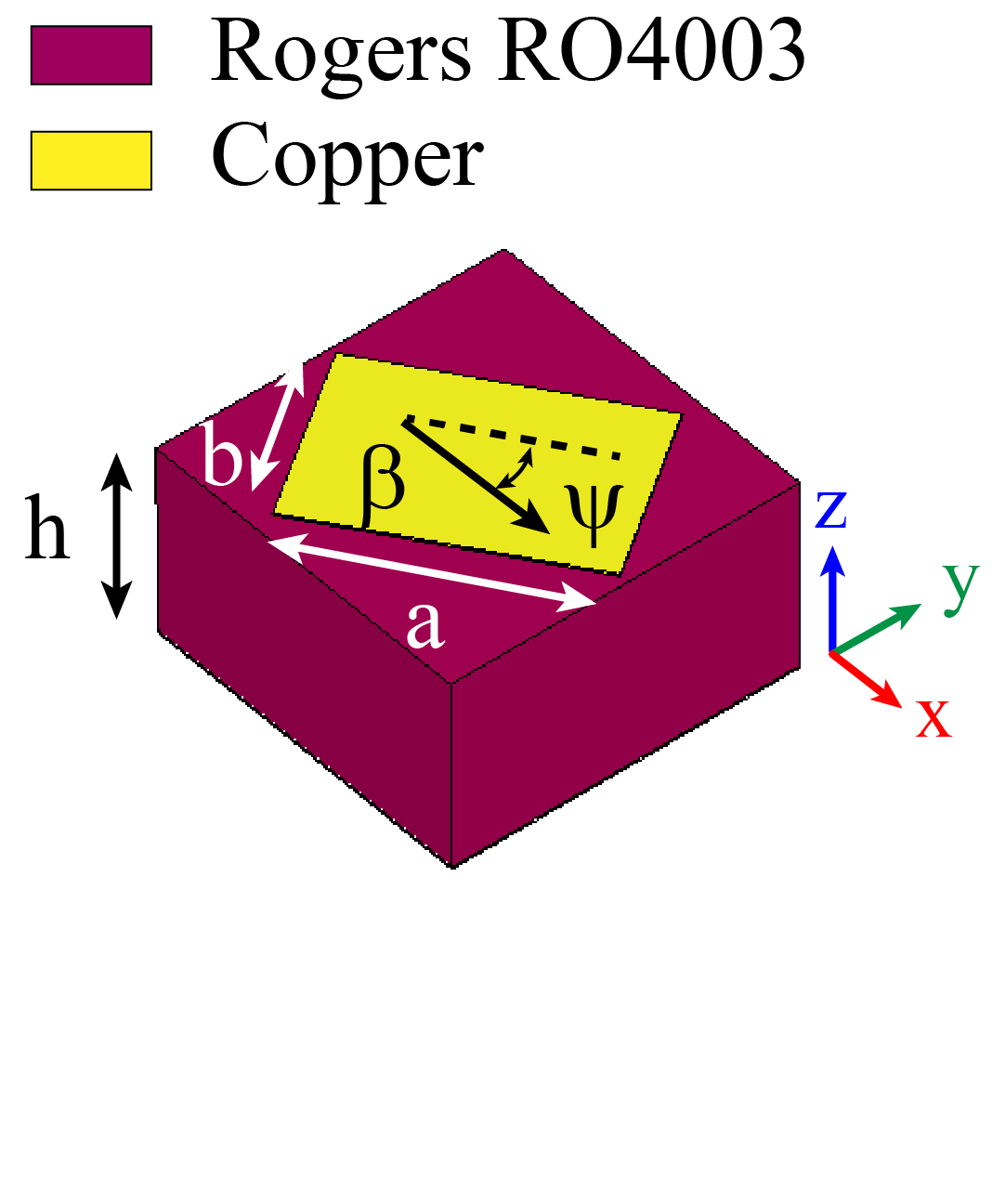}}
     %%%%%  
      \subfloat[\label{fig:Fig4b}]{%
       \includegraphics[width=0.27\textwidth]{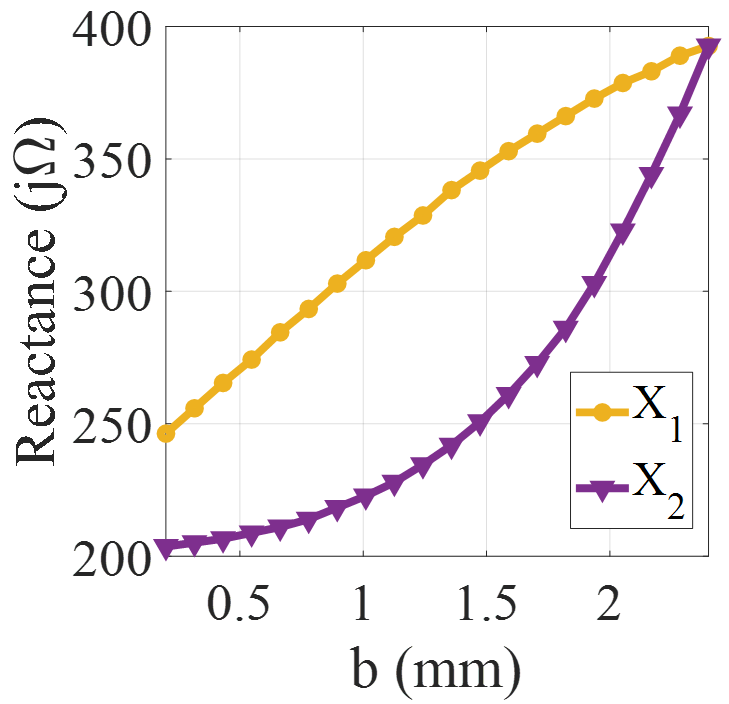}}
       \caption{(a) Proposed asymmetric unit-cell for the implementation of MoMetA. The parameter $\psi$ is the angle of rotation relative to the wave propagation, (b) impedance curves for variation of width of patch.}
\end{figure}

A method for the synthesis of anisotropic impedance surfaces is through using asymmetric patches printed on grounded dielectrics. Various shapes have been proposed in the literature. Elliptical \cite{mencagli2015},  rectangular \cite{amini2020}, slotted patches \cite{fong2010}, \cite{minatti2012}  and anchor-shaped elements \cite{faenzi2019} are among the most common geometries for realizing anisotropic impedance surfaces. Among the mentioned geometries, rectangular and elliptical structures have wider bandwidth and are less sensitive to the fabrication errors \cite{faenzi2019}. Meanwhile, these two geometries cover our required impedance range. 
In this paper, rectangular patches are used to synthesize the required surface impedance patterns as shown in Figs \ref{fig:Fig3a} and \ref{fig:Fig3b}. 
Fig.\ref{fig:Fig4a} shows the proposed unit-cell which possesses axial asymmetry to realize tensorial impedance. Rogers RO4003 with dielectric constant of 3.5, $\tan \delta$ of 0.0027 and thickness of 1.524 mm is used for the substrate. The unit-cell period is selected 2.8 mm, which is equal to $\lambda/6$ at the design frequency (18 GHz). Therefore, the 2-D periodic arrangement of this unit-cell can effectively model the impedance surface.
\begin{figure}
\centering
     \subfloat[\label{fig:Fig5a}]{%
       \includegraphics[width=0.25\textwidth]{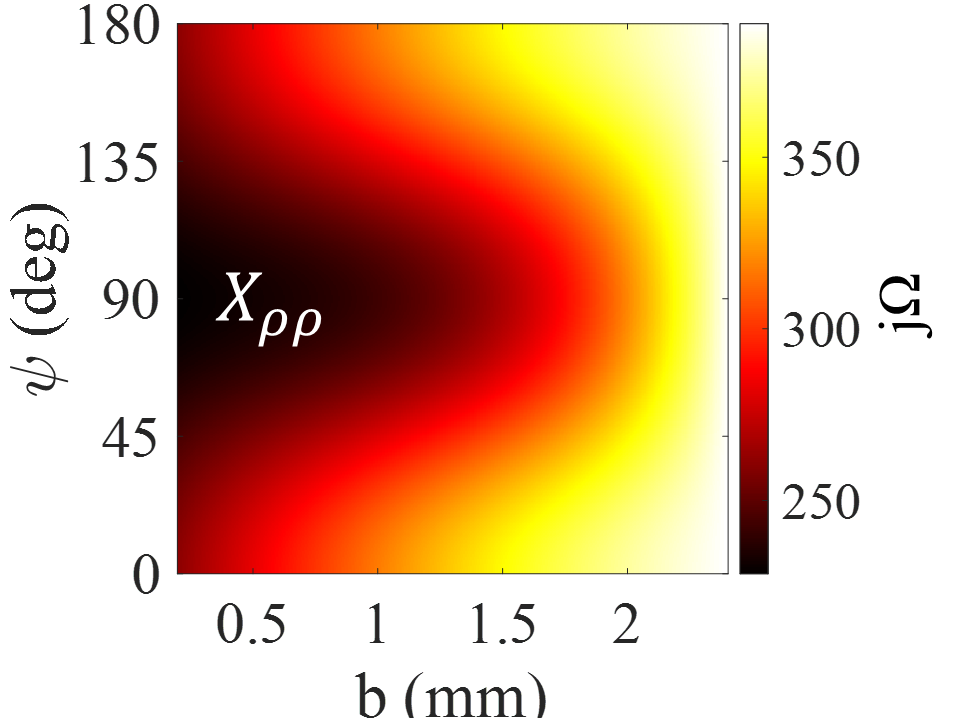}}
     %%%%%  
      \subfloat[\label{fig:Fig5b}]{%
       \includegraphics[width=0.25\textwidth]{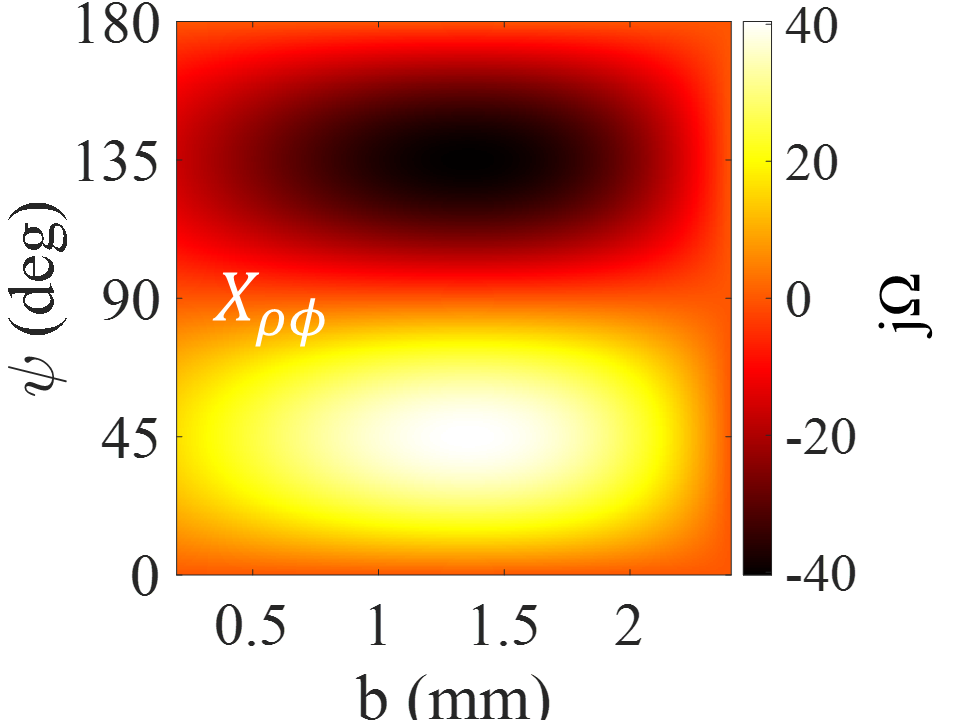}}
       \caption{Design maps for tensor components; (a) $X_{\rho\rho}$, (b) $X_{\rho\phi}$.}
\end{figure}
In Fig.\ref{fig:Fig4a} the parameter \textit{a} is kept constant and is equal to 2.4 mm. Two parameters of width of patch (namely \textit{b}) and rotation angle with respect to the propagation vector (namely $\psi$) are considered as variables. The proposed method in \cite{amini2020} and \cite{werner2014} is used to extract the surface impedance based on \textit{b} and $\psi$ variations. In Fig.\ref{fig:Fig4b} two reactance curves called $X_1$ and $X_2$ are extracted. Reactance $X_1$ is obtained for wave propagation along the x axis, while reactance $X_2$ is for that along the y axis. In both cases, the parameter $\psi$ is kept equal to zero and the parameter \textit{b} changes from 0.2 to 2.4 mm. Now the surface impedance matrix can be written based on the variations in \textit{b} and $\psi$:
\begin{equation}
\bar{\bar{Z}}_s = j\bar{\bar{X}}_s = jR^T(\psi)\bar{\bar{X}}(b)R(\psi)
\label{eq:jXs}
\end{equation}
where $\bar{\bar{X}}(b)$ is a diagonal matrix consisting of $X_1$ and $X_2$ as a function of \textit{b} (according to Fig.\ref{fig:Fig4b}) and is defined as:
\begin{equation}
\bar{\bar{X}}(b) = 
\begin{bmatrix}
X_1(b) & 0 \\
0 & X_2(b)\\
\end{bmatrix}
\end{equation}
and \textit{R} is the rotational matrix in Euclidean space:
\begin{equation}
R(\psi) = 
\begin{bmatrix}
\cos \psi & -\sin \psi \\
\sin \psi & \cos \psi\\
\end{bmatrix}
\end{equation}
Note that, the superscript \textit{T} indicates the transpose operator. Based on (\ref{eq:jXs}), the design maps are obtained as illustrated in Figs \ref{fig:Fig5a} and \ref{fig:Fig5b}.
Since $X_{\rho\rho}$ and $X_{\rho\phi}$ are determined anywhere on the hologram, then \textit{b} and $\psi$ for each unit-cell can be obtained by solving the non-linear equation system in (\ref{eq:jXs}) by the method of least squares. Fig.\ref{fig:Fig6} shows the required dimensions and rotation angles of rectangular patches related to the corresponding surface reactances in Figs \ref{fig:Fig3a} and \ref{fig:Fig3b}.
\begin{figure}
\centering
     \subfloat[\label{fig:Fig6a}]{%
       \includegraphics[width=0.25\textwidth]{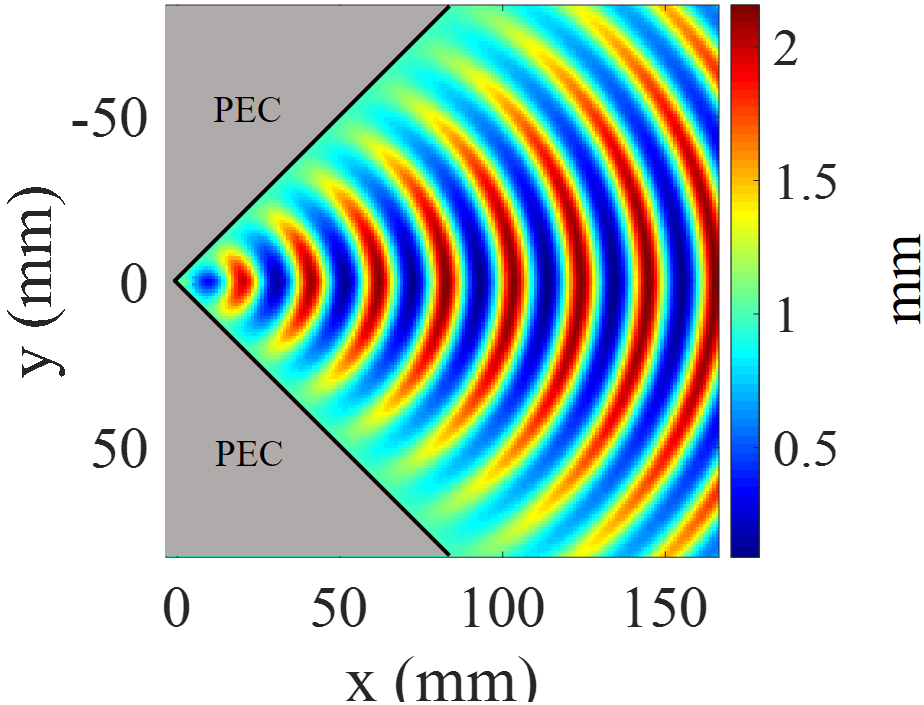}}
     %%%%%  
      \subfloat[\label{fig:Fig6b}]{%
       \includegraphics[width=0.25\textwidth]{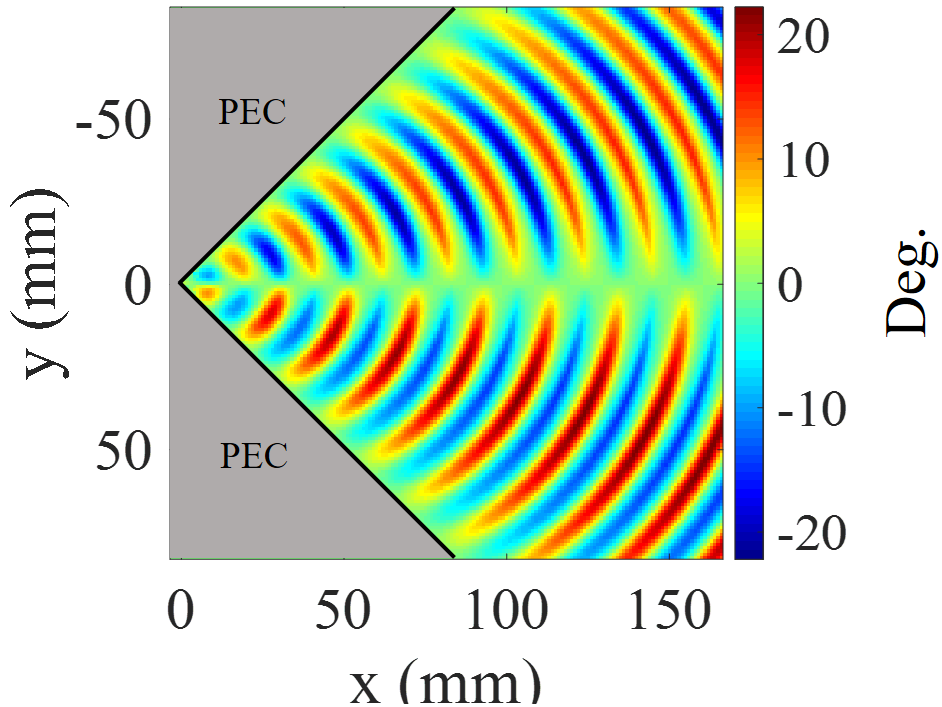}}
       \caption{Required values for (a) patch dimensions and (b) their angles for the implementation of surface reactances in Figs \ref{fig:Fig3c} and \ref{fig:Fig3d}.}
       \label{fig:Fig6}
\end{figure}
 \subsection{Implementation of surface wave reflector}
  In an ideal case a perfect electric conductor (PEC) wall with tall height is required for the realization of wedge-shaped reflector. This approach, however, is not suitable for integrated structures. Since leaky-wave MoMetAs are surface-wave-type structures and a noticeable portion of power is confined on their surface, the thickness of PEC wall can be considered thin, as long as it does not deteriorate the operation of antenna. 
 \begin{figure}
\includegraphics[width = 0.47\textwidth]{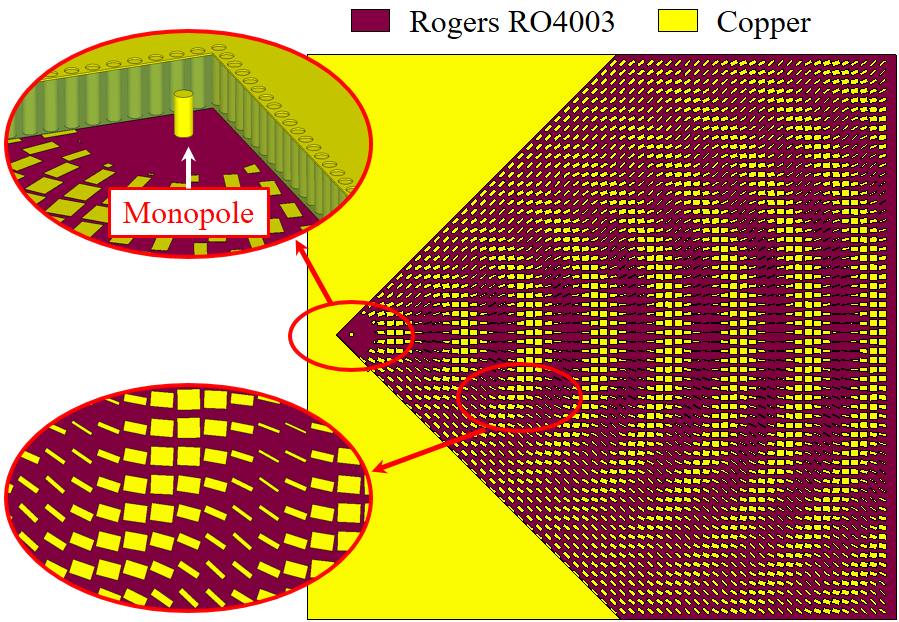}
\caption{Proposed reflector-based anisotropic MoMetA.}
\label{fig:antenna}
\end{figure}
In order to realize the PEC wall as the surface wave reflector, a row of metallized vias as shown in Fig.\ref{fig:antenna} are implemented on an FR-4 dielectric substrate with the thickness of 3.2 mm. The top and bottom surfaces of this substrate is covered by copper. The diameter of vias and the distances among them are chosen so that the row of vias acts as a metallic wall \cite{moeini_PRA2019} and prevents the wave leakage.
A monopole antenna is located at a distance of 5 mm from the corner of wedge-shaped reflector. In order to adjust the impedance bandwidth of hologram a circular region with radius of 7 mm is subtracted from the patches around the monopole. 
 \subsection{Simulation results of reflector-based MoMetA}
\begin{figure}
\centering
     \subfloat[\label{fig:scan_conventional}]{%
       \includegraphics[width=0.5\textwidth]{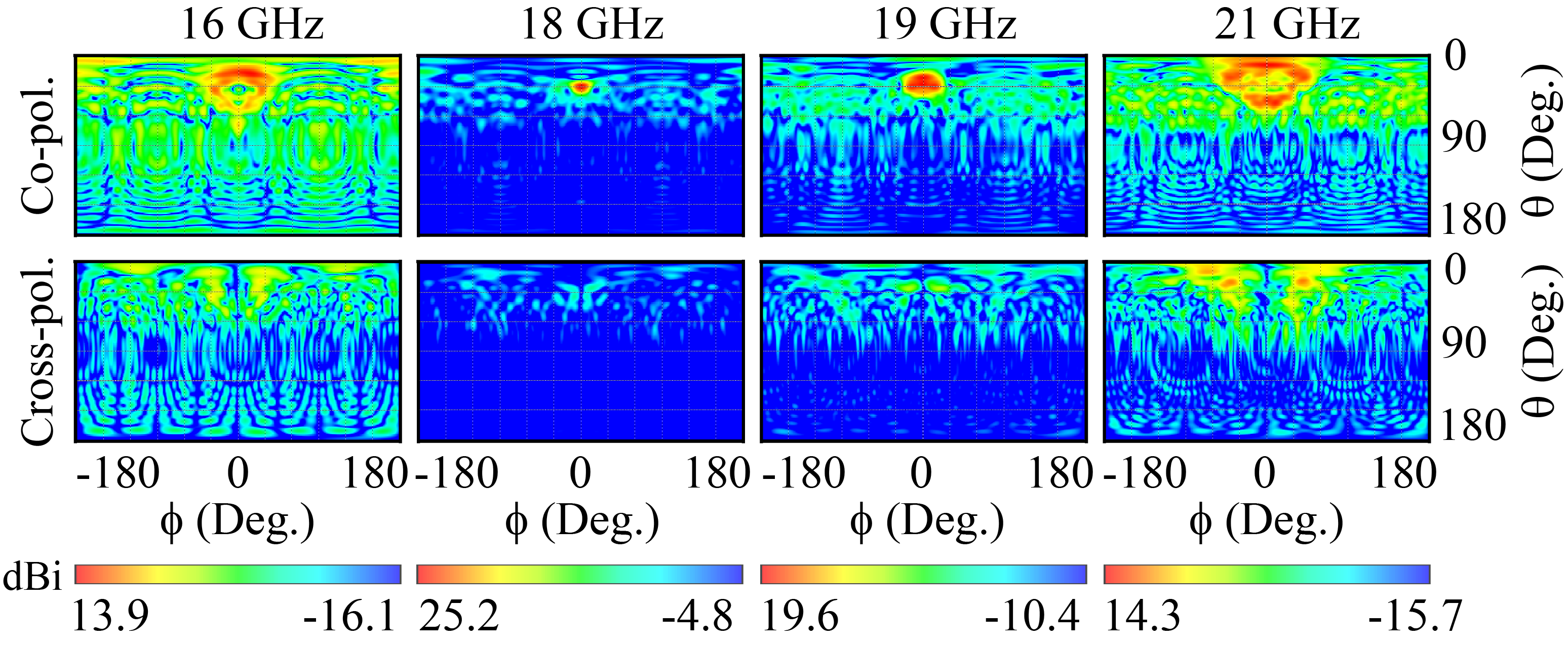}}\\
      \subfloat[\label{fig:scan_reflector}]{%
       \includegraphics[width=0.5\textwidth]{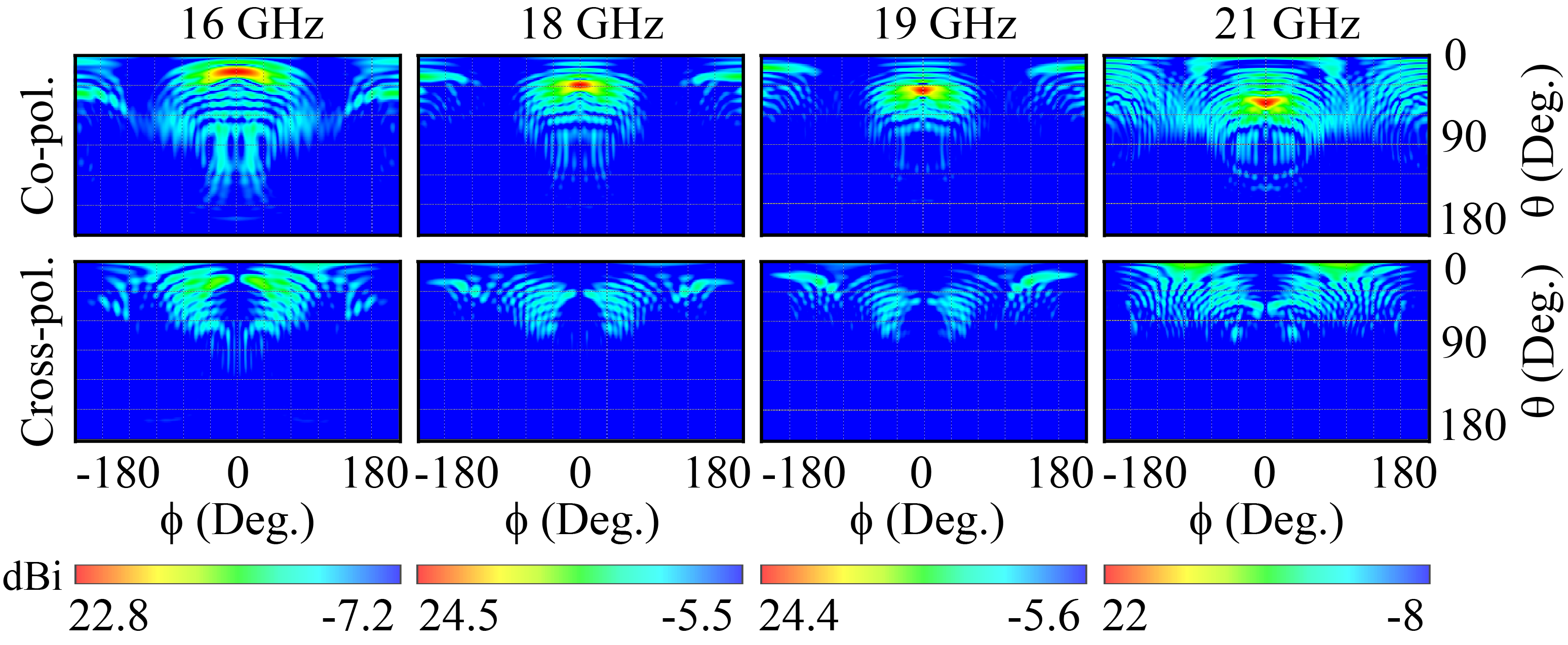}}
       \caption{Comparison between frequency scannability of (a) center-fed anisotropic MoMetA and (b) reflector-based anisotropic MoMetA.}
       \label{fig:scan_conv_ref}
\end{figure}

\begin{figure}
\includegraphics[width = 0.49\textwidth]{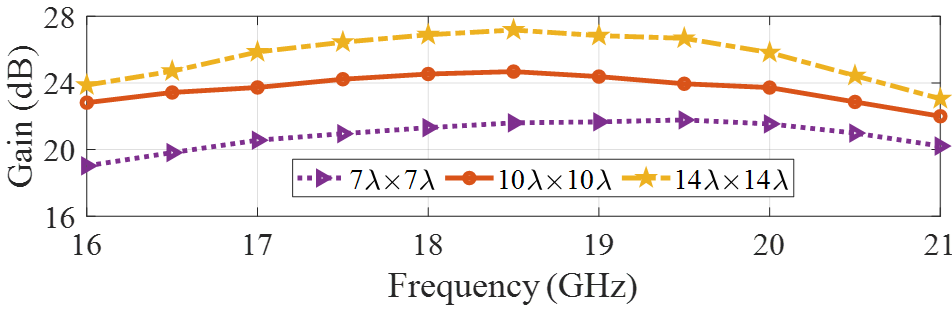}
\caption{Comparison of gain for different antenna dimensions.}
\label{fig:Gain_Comp}
\end{figure}

 Fig.\ref{fig:scan_conv_ref} shows the simulation results of radiation patterns for both cases of conventional MoMetA (center-fed anisotropic MoMetA) and reflector-based anisotropic MoMetA, respectively. The simulations are performed by CST Microwave Studio and transient analysis method \cite{CST}. The results indicate that using the surface wave reflector significantly increases the radiation bandwidth, since it eliminates the destructive effects of backward modes.  The dimensions of both holograms are equal to $10\lambda\times 10\lambda mm^2$. In the range of 16 to 21 GHz, the gain of reflector-based antenna varies  from 22 to 24.5 dB. Also the side-lobe-level (SLL) variation in the scan plane are obtained from -13.8 to -12.5 dB, which is desired. The key point here is the suitable cross-polarization level, where the co- to cross-polarization level ratio varies from 38.2 to 44.4 dB. The details of simulation results at other frequencies are listed in Table \ref{tab:scan_conv_ref}.
The gain of co-pol component depends on the dimensions of the effective radiation area. Increasing the dimensions of the aperture (radiation area) will enhance the co-pol gain.
In Fig.\ref{fig:Gain_Comp}  antenna gains for different dimensions are compared.
\section{Design of polarization converter}
 In this section, the principles of polarization converter performance and its design procedure are discussed.  For the analysis and design, the polarizer is considered to be illuminated by an ideal plane wave. After designing the polarizer with an acceptable angular stability, it will be fixed over the holographic antenna.
 \begin{table}
\caption{Comparison of anisotropic reflector-based MoMetA with conventional center-fed MoMetA.}
\label{tab:my-table}
\begin{tabular}{c|ccc|ccc}
\hline\hline
                                                      & \multicolumn{3}{c|}{Center-fed MoMetA}                                                                                                                                         & \multicolumn{3}{c}{Reflector-based MoMetA}                                                                                                                                     \\ \hline
\begin{tabular}[c]{@{}c@{}}Freq.\\ (GHz)\end{tabular} & \begin{tabular}[c]{@{}c@{}}Co-pol.\\ (dB)\end{tabular} & \begin{tabular}[c]{@{}c@{}}Cross-pol.\\ (dB)\end{tabular} & \begin{tabular}[c]{@{}c@{}}SLL\\ (dB)\end{tabular} & \begin{tabular}[c]{@{}c@{}}Co-pol.\\ (dB)\end{tabular} & \begin{tabular}[c]{@{}c@{}}Cross-pol.\\ (dB)\end{tabular} & \begin{tabular}[c]{@{}c@{}}SLL\\ (dB)\end{tabular} \\ \hline
16                                                    & 13.9                                                   & -10.3                                                     & -4.9                                               & 22.8                                                   & -21.1                                                     & -13.8                                              \\
17                                                    & 17.6                                                   & -6.0                                                      & -3.5                                               & 23.7                                                   & -20.7                                                     & -13.7                                              \\
18                                                    & 25.2                                                   & -7.9                                                      & -18.6                                              & 24.5                                                   & -17.7                                                     & -13.3                                              \\
19                                                    & 19.6                                                   & -5.6                                                      & -13.4                                              & 24.4                                                   & -16.8                                                     & -13.4                                              \\
20                                                    & 15.9                                                   & -8.1                                                      & -2.3                                               & 23.7                                                   & -18.6                                                     & -13.5                                              \\
21                                                    & 14.3                                                   & -4.5                                                      & -1.6                                               & 22                                                     & -16.2                                                     & -12.5                                              \\ \hline
\end{tabular}
\label{tab:scan_conv_ref}
\end{table}
 \subsection{Principle of operation}
\begin{figure}
  \includegraphics[width = 0.45\textwidth]{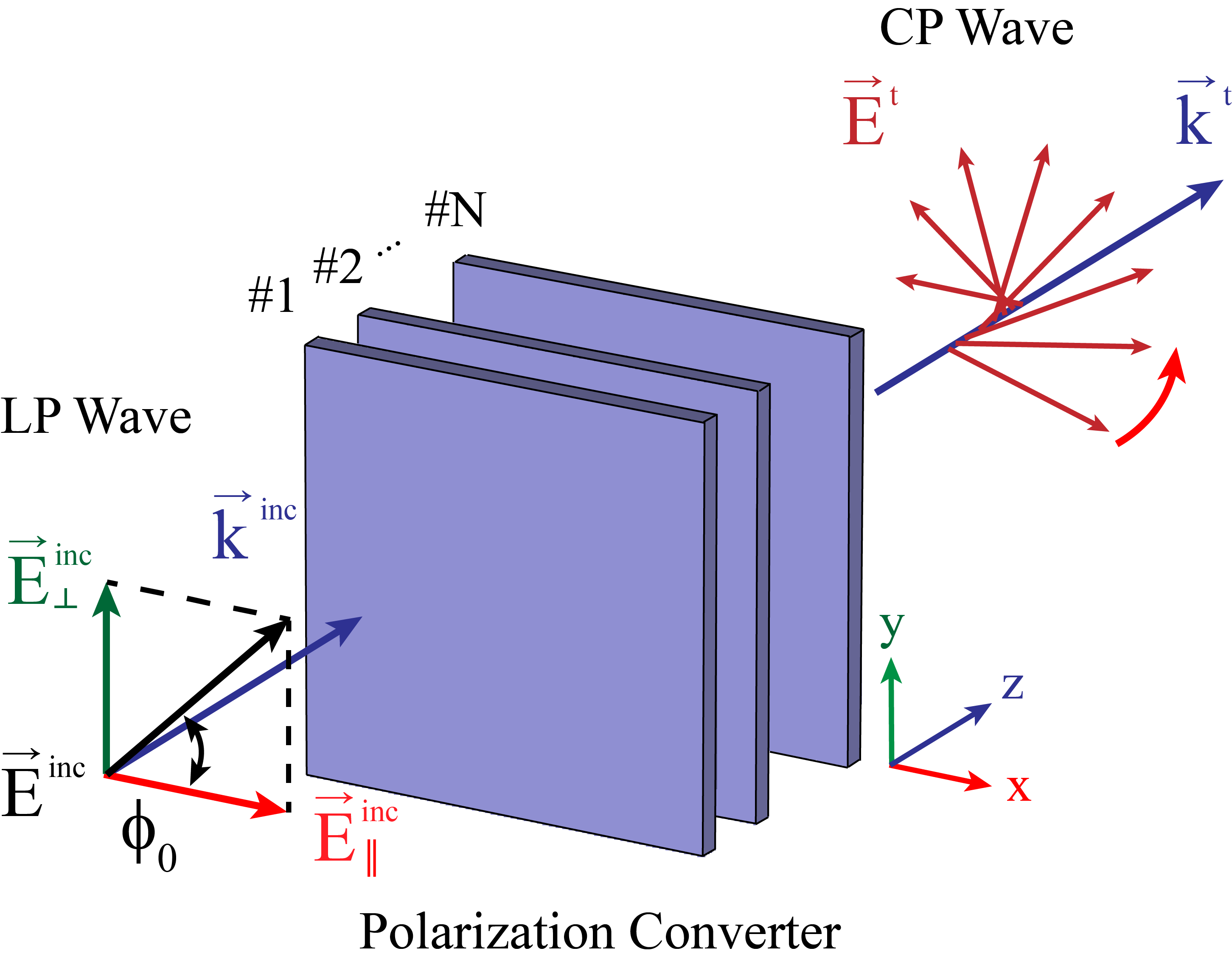}
  \caption{Conceptual geometry of a multi-layer linear to  circular polarization converter.}
  \label{fig:Polarizer}
  \end{figure}

  Fig.\ref{fig:Polarizer} shows the conceptual topology of a linear to circular polarization converter, which consists of \textit{N} layers separated by air gaps.  The structure is excited by an incident plane wave with linear polarization.
The direction of electric field of incident wave, $\vec{E}^{inc}$, has an angle of $\phi_0 = 45^\circ$ with the x axis.   
The direction of wave propagation is perpendicular to the structure ( z direction as shown in Fig.\ref{fig:Polarizer}).
Therefore, the incident electric field can be decomposed into two components of parallel (x pol.) and perpendicular (y pol.) to the x axis. Note that the two components have equal amplitudes.
  Therefore, we have:
   \begin{equation}
 \vec{E}^{inc} = \vec{E}^{inc}_\perp + \vec{E}^{inc}_\parallel
 \label{eq:Einc}
 \end{equation}
 This polarization converter acts differently towards each electric field component. Inside the operational frequency bandwidth, the amplitude of transmission coefficient is approximately equal for both components and the structure transmits both components with no loss. Meanwhile, parallel and perpendicular components each experiences different phase shifts, and their phase difference after passing through the designed structure will be 90 degrees. As a result, the emerging wave polarization on the other side of structure will be circular. The transmitted wave at the output of polarizer may be written as a summation of two linear parallel and perpendicular waves ($\vec{E}^t_\perp$ and $\vec{E}^t_\parallel$):
  \begin{equation}
 \vec{E}^t = \vec{E}^t_\perp + \vec{E}^t_\parallel = E_0 (T_\perp \hat{a}_x + T_\parallel \hat{a}_y) e^{-jkz}
 \end{equation}
 where $T_\perp = |T_\perp|e^{j\angle T_\perp}$ and $T_\parallel = |T_\parallel|e^{j\angle T_\parallel}$ are transmission coefficients of designed polarizer for perpendicular and parallel polarizations, respectively. In order to achieve a circularly polarized wave at the output of polarizer inside the frequency bandwidth, the following relations between $T_\perp$ and $T_\parallel$ should be applied:
 \begin{equation}
 |T_\perp| = |T_\parallel|
 \end{equation}
 \begin{equation}
 \angle T_\perp - \angle T_\parallel = \pm \frac{\pi}{2}
 \label{eq:angle}
 \end{equation}
 The "$\pm$" sign in (\ref{eq:angle}) determines the direction of wave rotation with output circular polarization. The "+" sign is related to the transmitted wave with left hand polarization, while the "-" sign is related to the right hand one. The common standard for demonstrating the quality of circular polarization is the axial ratio (AR), obtained as \cite{lopez2014}:
 \begin{equation}
 AR = (\frac{|T_\perp|^2 + |T_\parallel|^2 + \sqrt{a}}{|T_\perp|^2 + |T_\parallel|^2 - \sqrt{a}})^{1/2}
 \label{eq:AR}
\end{equation}  
\begin{equation}
a = |T_\perp|^4 + |T_\parallel|^4 + 2 |T_\perp|^2|T_\parallel|^2 \cos (2(\angle T_\parallel - \angle T_\perp))
\label{eq:a}
\end{equation}
\begin{figure}
\centering
\subfloat[\label{fig:UC}]{%
       \includegraphics[width=0.45\textwidth]{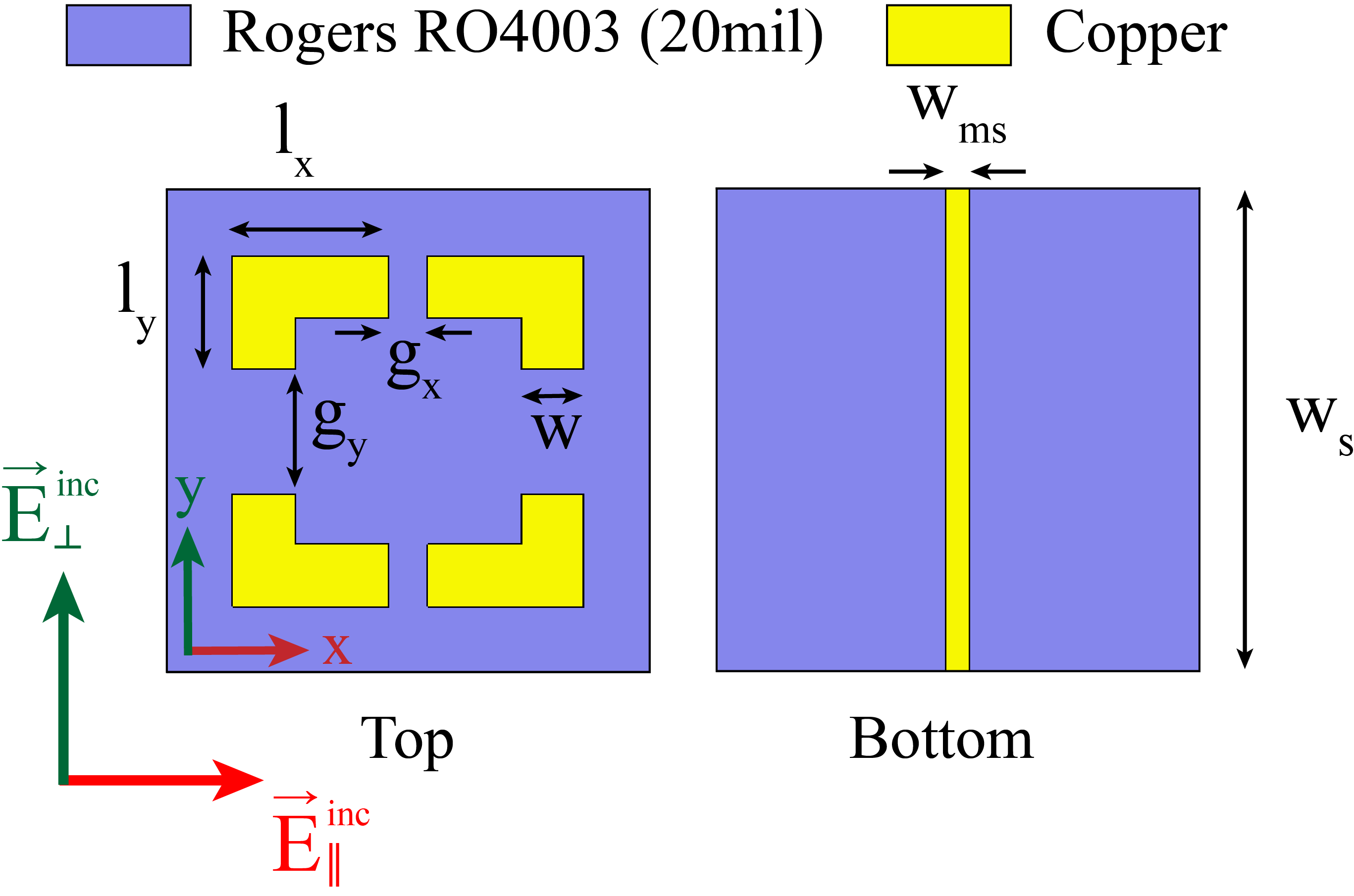}}\\
\subfloat[\label{fig:EQ}]{%
       \includegraphics[width=0.43\textwidth]{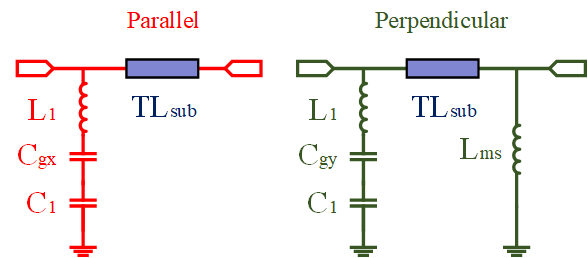}}
\caption{(a) Proposed unit-cell for realization of polarization converter, (b) equivalent circuit models for parallel and perpendicular incident wave.}
\label{fig:UC_EQ}
\end{figure}
\subsection{Unit-cell design}
In order to design a polarizer suitable for placing over the holographic antenna, the following specifications should be considered:
\begin{enumerate}
\item The polarizer should have minimum transmission loss.
\item The polarizer should have the specified bandwidth.
\item The polarizer should have minimum sensitivity to the incident wave angle. Here, it should have angular stability in the range of 20 to 47 degrees in elevation.
\end{enumerate}
Fig.\ref{fig:UC} shows the proposed unit-cell of polarizer. Its structure is based on the unit-cell introduced in \cite{lopez2014} and \cite{youn2018}. Our designed structure has a fewer number of layers and wider range of angular stability in comparison with the proposed unit-cells in \cite{lopez2014} and \cite{youn2018}. By increasing the number of layers, it would be possible to achieve even wider bandwidths. Nevertheless, due to the limited bandwidth of antenna, from 16 to 21 GHz, adding polarizer layers only increases the final cost while it has no effect on the overall performance of antenna. The proposed unit-cell consists of three metasurface layers, each of which is made of Rogers RO4003 with thickness of 0.508 mm. Each three layers contains the same and equal number of unit-cells with an air gap distance of \textit{d}. The main feature of square ring structure is its reflection resonance behavior \cite{lopez2014}. By adding gaps and narrow metal strips to the square ring, different transmission characteristics will happen for the two parallel and perpendicular polarizations. Substituting air gaps instead of dielectric layers will also lead to lower fabrication cost, lower structure weight and lower transmission loss.
In order to explain the polarizer operation, we use the equivalent circuits in Fig.\ref{fig:EQ}.  Due to the asymmetry of  structure, the equivalent circuit for the parallel and perpendicular polarizations are different. The resonance circuit consisting of $L_1$ and $C_1$, represents the simple rectangular ring. The $C_{gx}$ and $C_{gy}$ capacitors indicate the gaps on the arms of ring. Also, $L_{ms}$ indicates the inductance of narrow metal strip in the unit-cell, which is parallel to the y axis. Note that, $L_{ms}$  appears only for the incident polarization parallel to the narrow metal strip (y pol.) and the strip is transparent to the incident wave for x polarization. In addition, $TL_{sub}$ represents the transmission line, which models the propagation of wave in the dielectric substrate. The proposed polarizer shows a capacitive behavior towards the incident parallel mode, $\vec{E}_\parallel^{inc}$, due to the narrow gap capacitive properties. While, in the case of perpendicular electric field, $\vec{E}_\perp^{inc}$, an inductive behavior is observed mainly due to the presence of narrow metal strips at the bottom face of dielectric.
The final model of equivalent circuit can be obtained by cascading the three equivalent circuits. The defined unit-cell in CST Microwave Studio software is optimized to get the desired reflection coefficient in the operational frequency bandwidth of holographic antenna. Note that in the optimization of capacitive gaps and the metal strip, the fabrication limits have been considered. The final parameters of full-wave optimizations are tabulated in Table \ref{tab:dims}.
\begin{table*}[]
\caption{Optimal dimensions of unit-cell parameters.}
\centering
\begin{tabular}{cccccccc}
\hline\hline
$w_s$ (mm) & $w_{ms}$ (mm) & $l_x$ (mm) & $l_y$ (mm) & $g_x$  (mm) & $g_y$ (mm) & $w$ (mm) & $h$ (mm) \\ \hline 
6.2    & 0.3    & 2    & 1.5     & 0.5   & 0.5     & 1.6     & 0.508    \\ \hline
\end{tabular}
\label{tab:dims}
\end{table*}
\begin{figure}
\subfloat[\label{fig:T}]{%
       \includegraphics[width=0.5\textwidth]{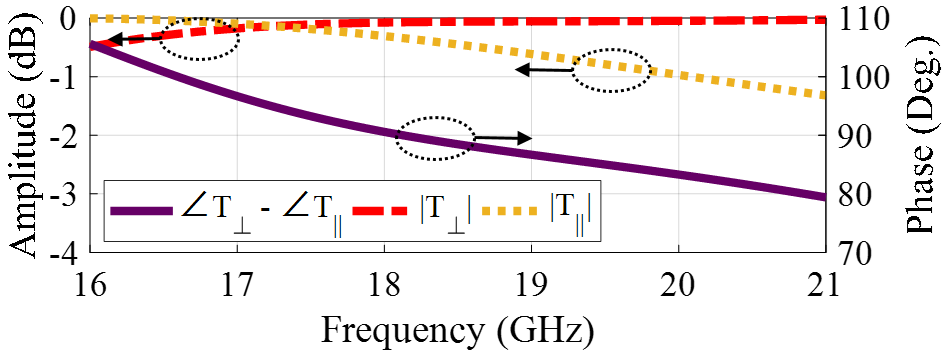}}\\
\subfloat[\label{fig:AR}]{%
       \includegraphics[width=0.45\textwidth]{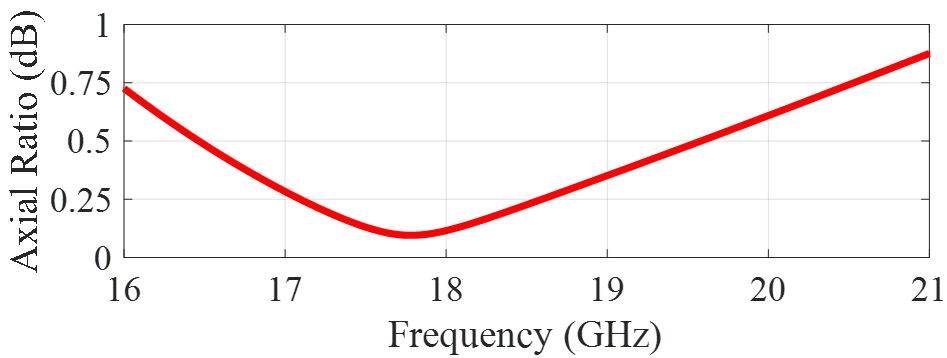}}
\caption{Frequency responses of the proposed unit-cell in Fig.\ref{fig:UC_EQ}: (a) The amplitude and phase of transmission coefficient, (b)  Axial ratio.}
\label{fig:T_AAR}
\end{figure}
Fig.\ref{fig:T_AAR} shows the simulated transmission coefficients  ($T_\perp, T_\parallel$) and axial ratio. The amplitude of  transmission coefficients in the range of 16 to 21 GHz are greater than -1.3 dB. Note that these results are obtained for the incident angle of 0 degrees. The phase difference in the range of 16 to 21 GHz is equal to $\delta \phi = 90^\circ \pm 15^\circ$, and the related axial ratio for this spectrum is lower than 1 dB. Fig. \ref{fig:rotation} demonstrates the electric field distribution on the other side of polarizer unit-cell for $\omega t = 0^\circ$, $\omega t = 90^\circ$, $\omega t = 180^\circ$ and $\omega t = 270^\circ$ at operational frequency of 18 GHz. Observe the rotation of electric field in one period in this figure. Note that, in order to observe the field rotation, the incident wave must contain both parallel and perpendicular components.
\begin{figure}
\centering
\includegraphics[width = 0.35\textwidth]{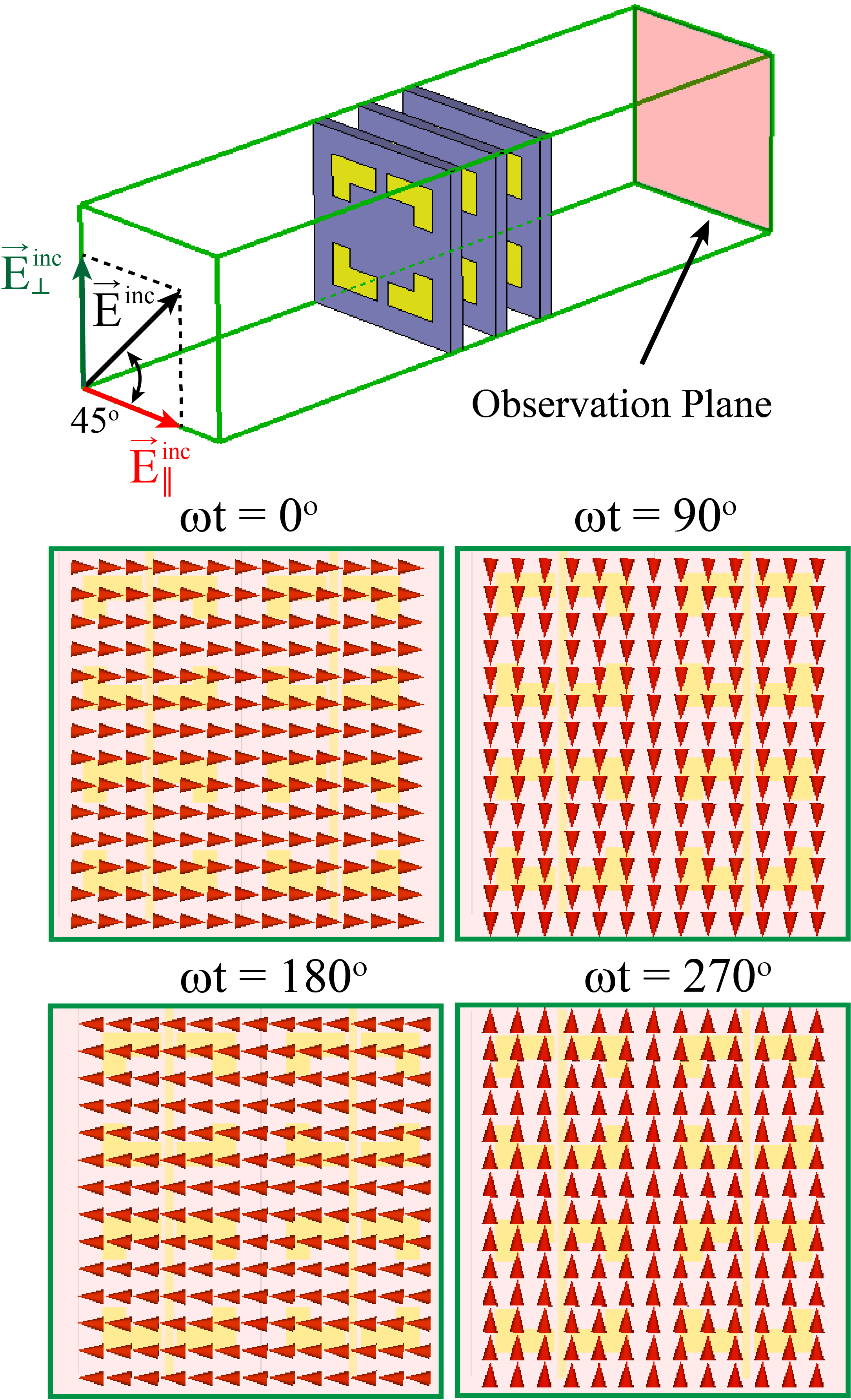}
\caption{Demonstration of electric field rotation at the other side of polrization converter.}
\label{fig:rotation}
\end{figure}
In this section, it is desired that the unit-cell experience the exact same illumination from the antenna designed in the previous section. This illumination should be plane wave with vertical polarization and the incident angle is specified from 20 to 47 degrees. 
\begin{figure}
\centering
\subfloat[\label{fig:T2}]{%
       \includegraphics[width=0.5\textwidth]{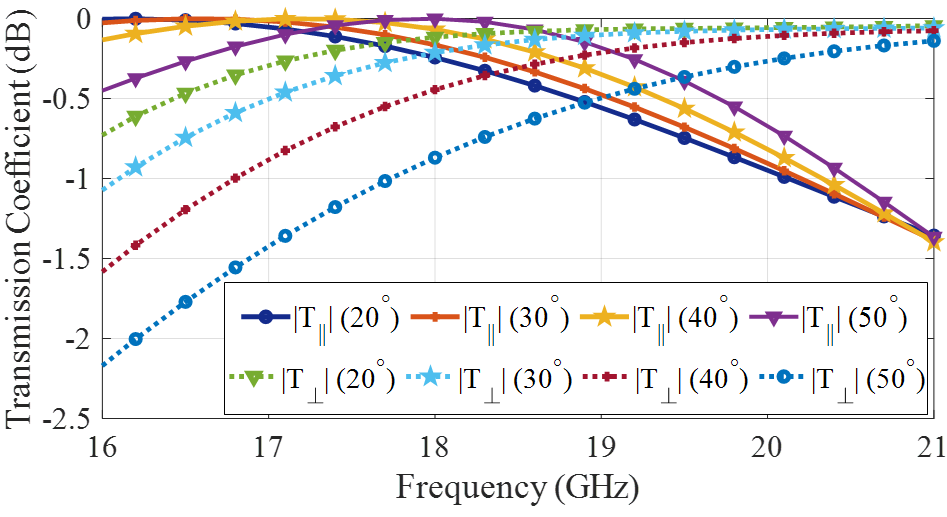}}\\
\subfloat[\label{fig:AR2}]{%
       \includegraphics[width=0.5\textwidth]{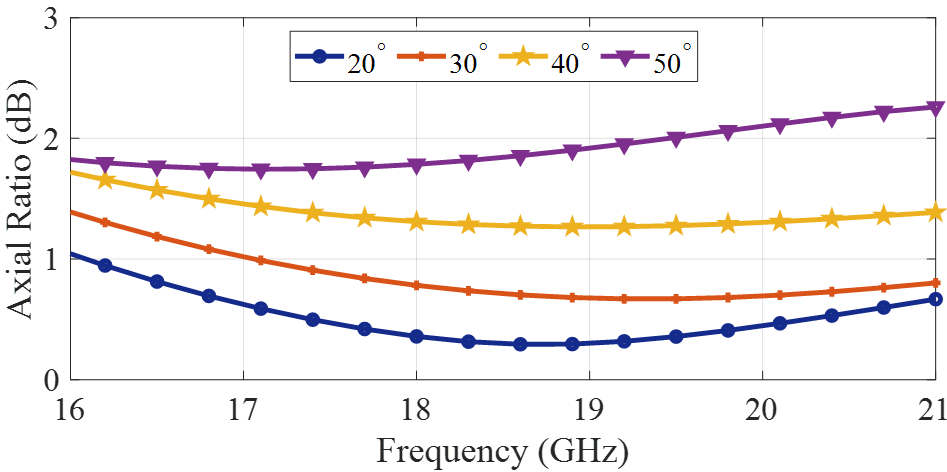}}\\
\subfloat[\label{fig:PER}]{%
       \includegraphics[width=0.5\textwidth]{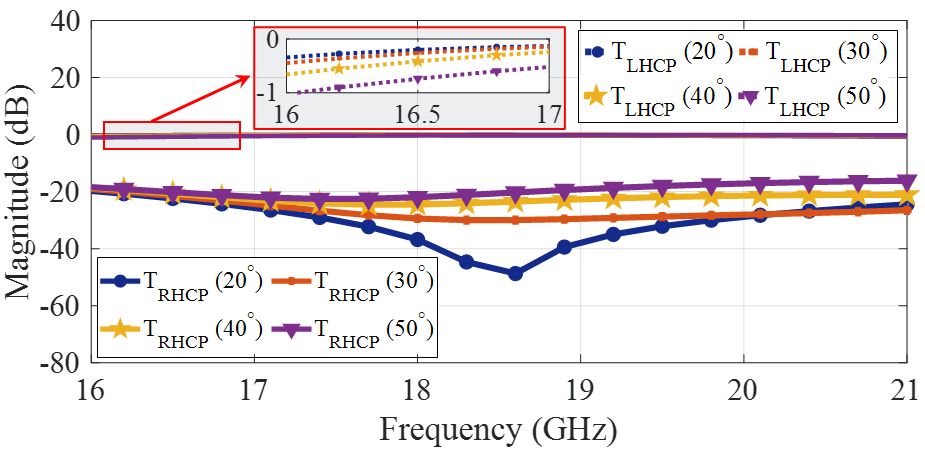}}   
\caption{Simulation results of the proposed unit-cell in Fig.\ref{fig:UC_EQ} for different incident angles: (a) amplitude of transmission coefficient, (b) axial ratio, (c) the magnitudes of LHCP, RHCP.}
\label{fig:ُT_AR2}
\end{figure}
Figs \ref{fig:T2} and \ref{fig:AR2} show the transmission coefficients and axial ratios for four incident angles between 20 to 50 degrees. Observe that the designed polarizer has an acceptable operation considering the amplitude of transmission signals. Also, in the frequency bandwidth of 16 to 21 GHz the axial ratio is less than 3dB for all of the incident angles, providing an acceptable angular stability.  To investigate the polarization purity of output radiation, magnitudes of the LHCP and RHCP components of transmitted wave are extracted, which are depicted in Fig.\ref{fig:PER}. The definitions of LHCP and RHCP components are given in more detail in \cite{yan2013}. Results show that the LHCP components are greater than -1.5 dB from 16 to 21 GHz for all incident angles. 
\section{Simulations and measurements of proposed antenna with the designed polarizer}
\begin{figure}
\includegraphics[width = 0.48\textwidth]{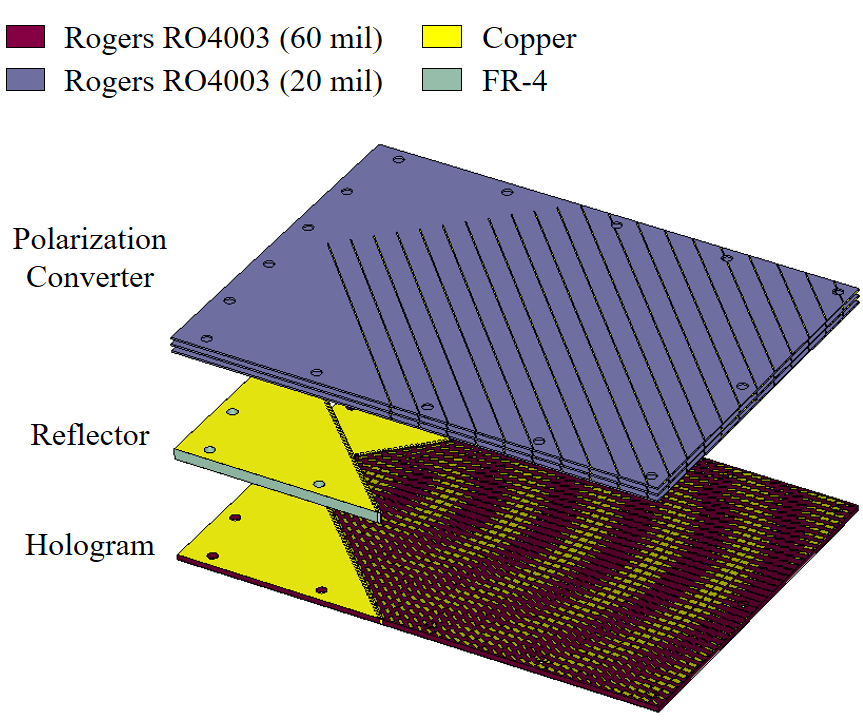}
\caption{The circularly polarized leaky-wave antenna consisting of reflector-based MoMetA and the polarization converter.}
\label{fig:Antenna_Exploded}
\end{figure}

\begin{figure*}
\centering
\includegraphics[width = 0.8\textwidth]{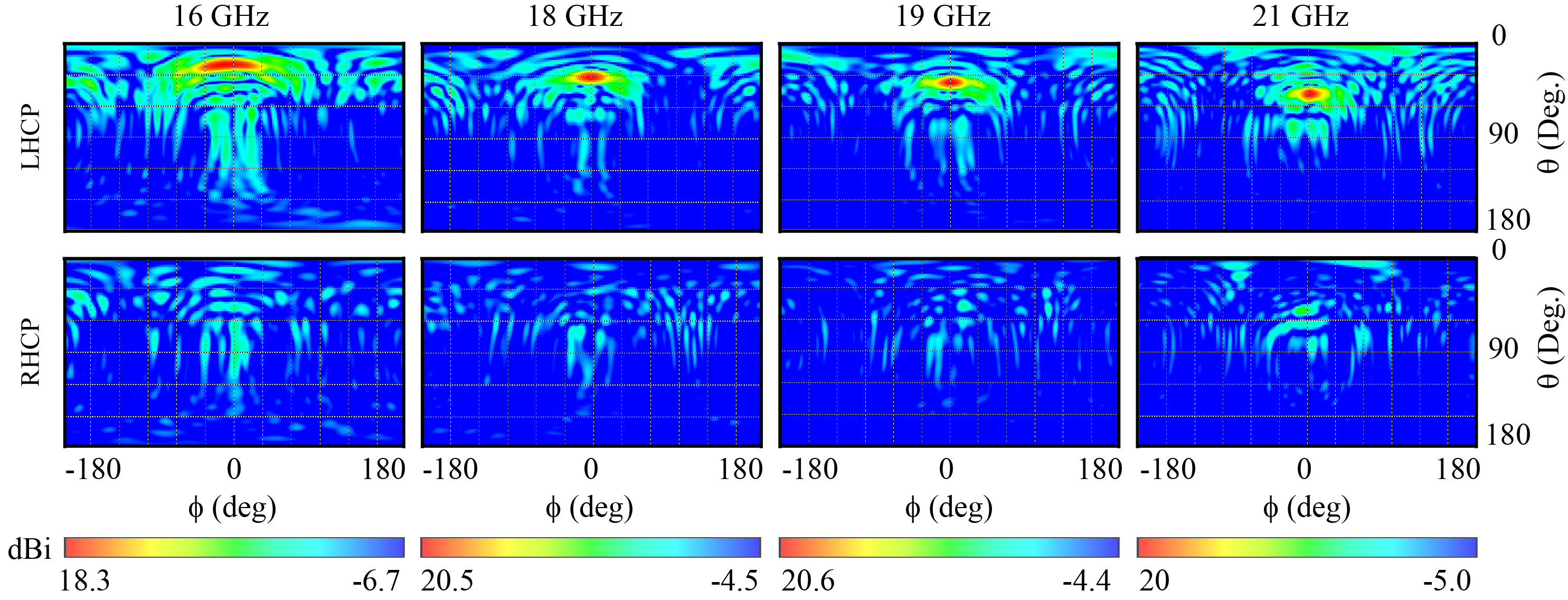}
\caption{Far-field patterns of reflector-based MoMetA in Fig.\ref{fig:Antenna_Exploded} at different frequencies.}
\label{fig:Pattern_Scan_Circ}
\end{figure*}

\begin{figure*}
\centering
\subfloat[\label{fig:Prot}]{%
       \includegraphics[width=0.298\textwidth]{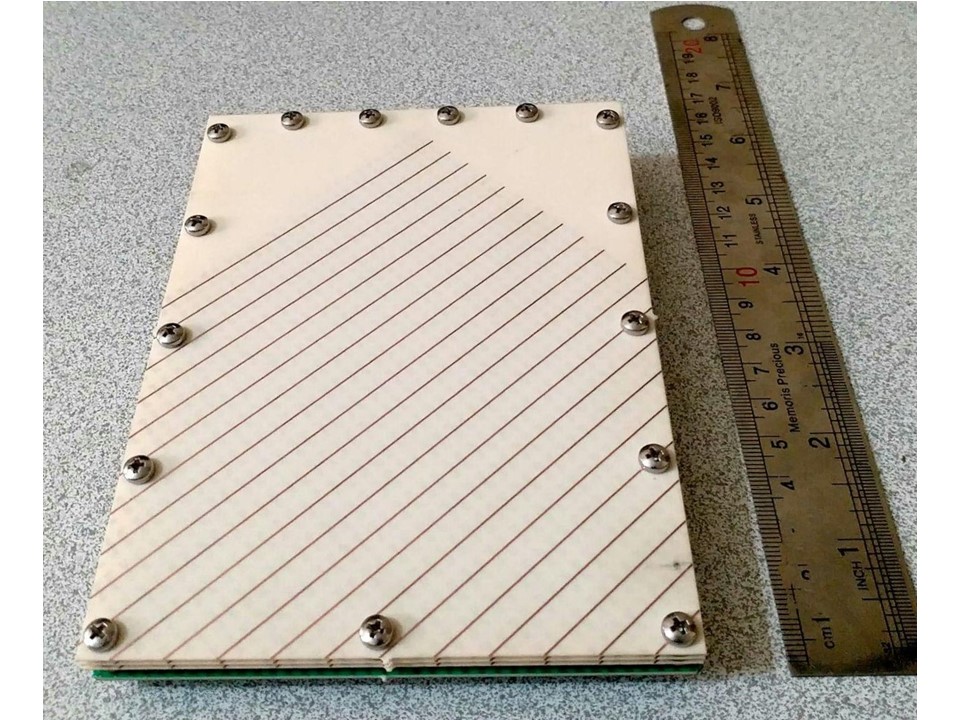}}
\subfloat[\label{fig:Prot2}]{%
       \includegraphics[width=0.32\textwidth]{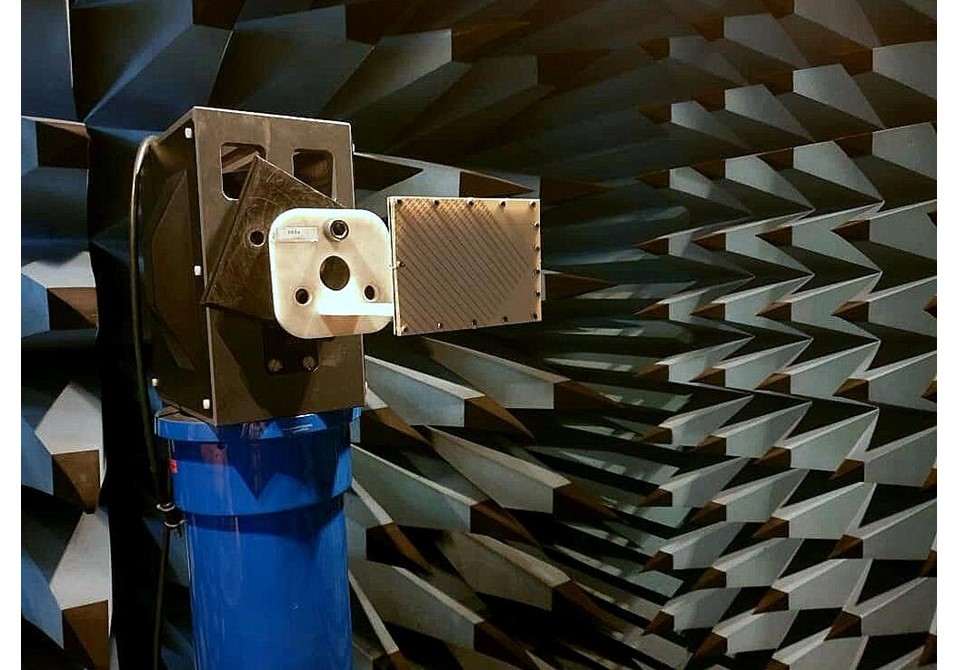}}
\subfloat[\label{fig:S11_Meas}]{%
       \includegraphics[width=0.3\textwidth]{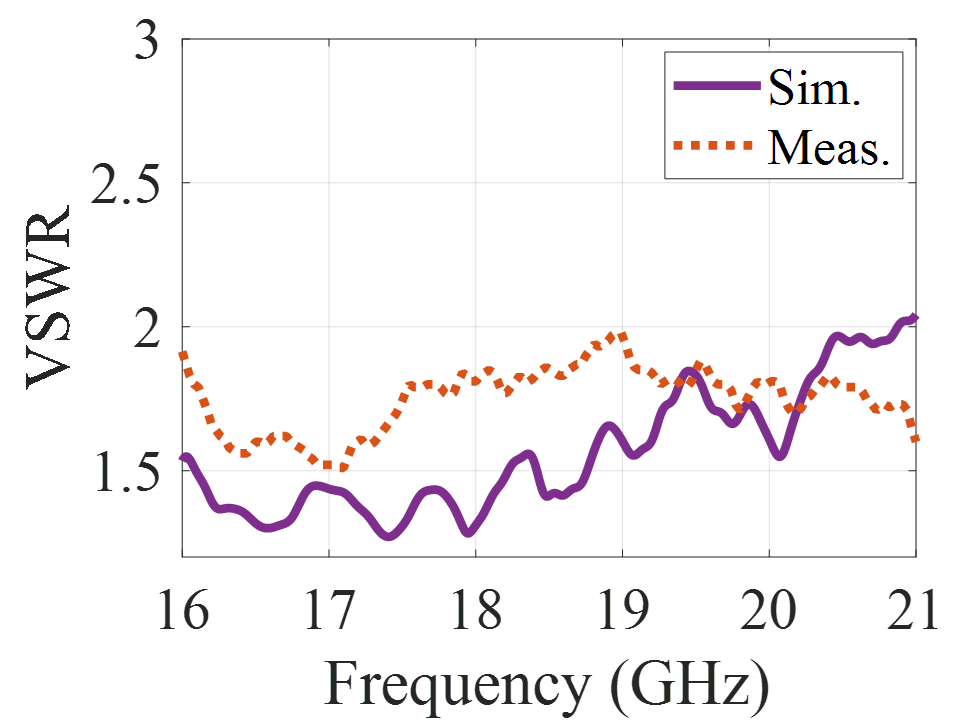}}   
\caption{(a) The prototype of reflector-based MoMetA with polarizer, (b) measurement setup, (c) simulation and measurement results of VSWR.}
\label{fig:ُProt_S11}
\end{figure*}

\begin{figure*}
\centering
\subfloat[\label{fig:Co16}]{%
       \includegraphics[width=0.25\textwidth]{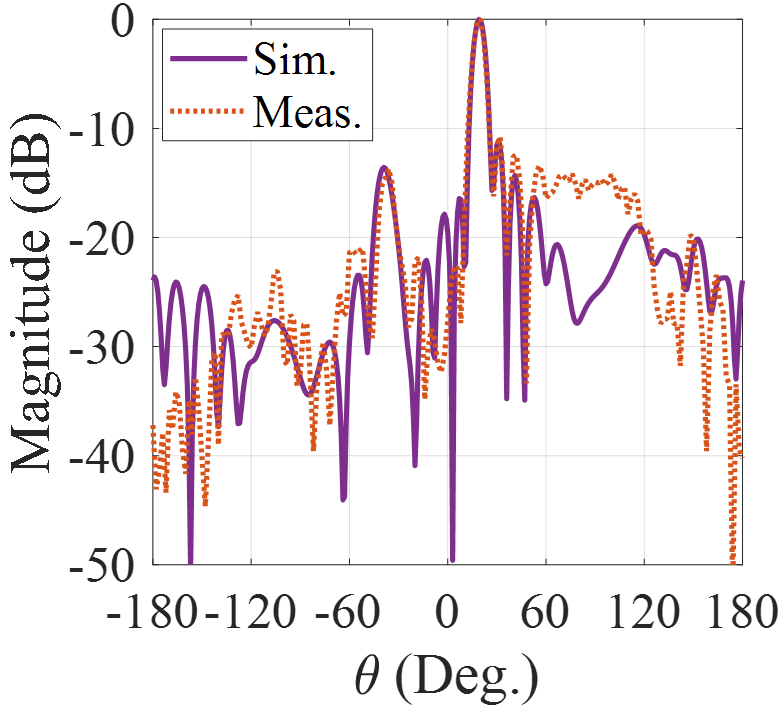}}
%\subfloat[\label{fig:Co17}]{%
%       \includegraphics[width=0.25\textwidth]{Co17.png}}
\subfloat[\label{fig:Co18}]{%
       \includegraphics[width=0.25\textwidth]{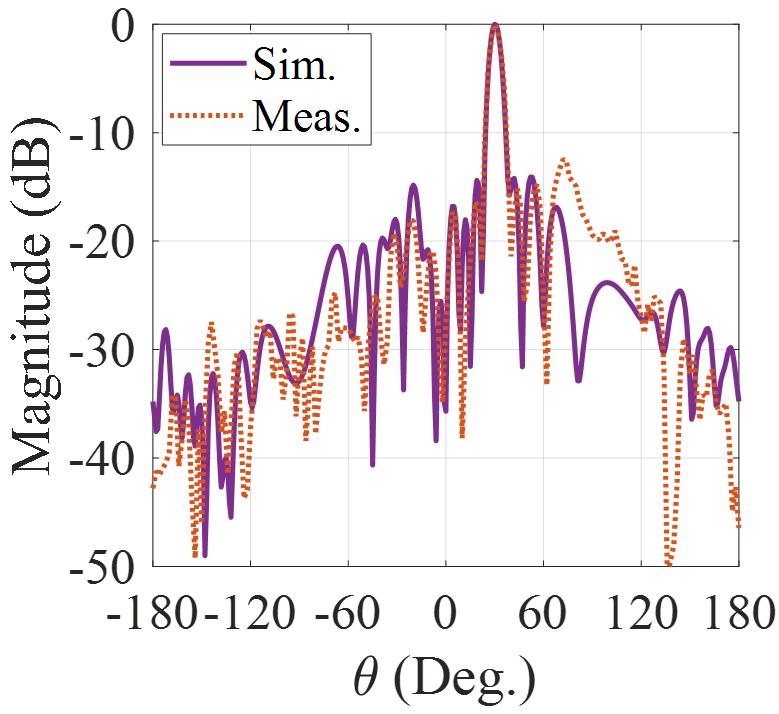}}
\subfloat[\label{fig:Co19}]{%
       \includegraphics[width=0.25\textwidth]{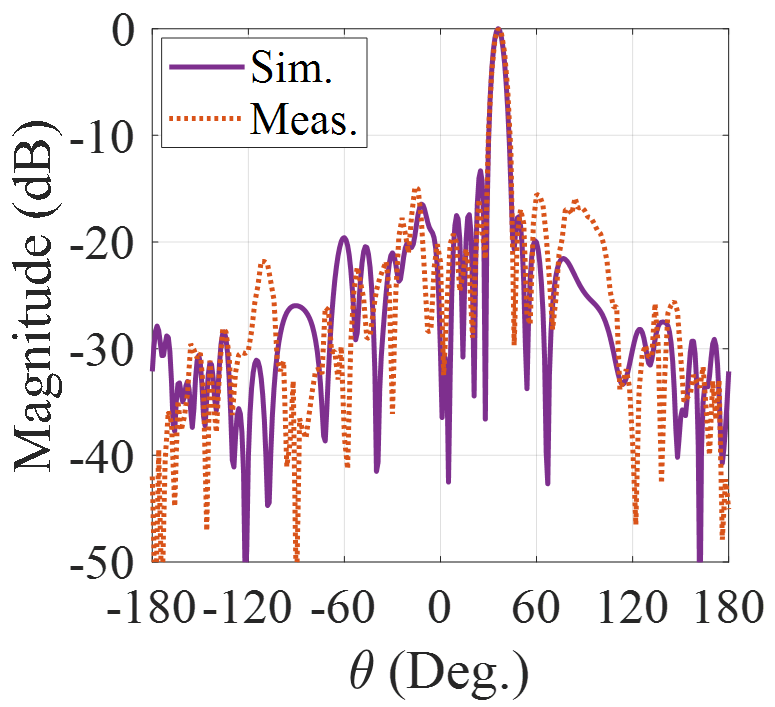}}
%\subfloat[\label{fig:Co20}]{%
 %      \includegraphics[width=0.25\textwidth]{Co20.png}}
\subfloat[\label{fig:Co21}]{%
       \includegraphics[width=0.25\textwidth]{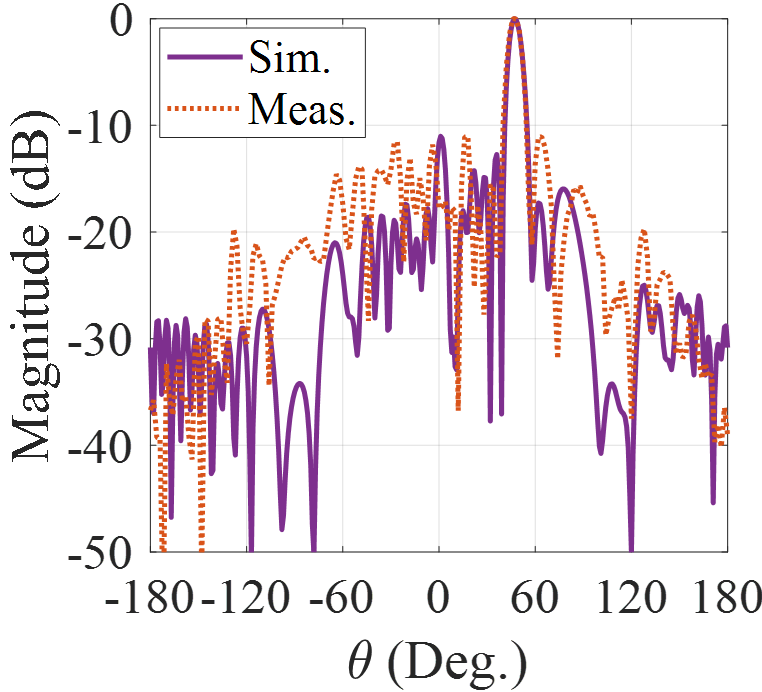}}
\caption{Comparison between simulation and measurement results of LHCP component of far-field patterns at (a) 16 GHz, (b) 18 GHz, (c) 19 GHz and (d) 21 GHz.}
\label{fig:Co_MEAS}
\end{figure*}

\begin{figure*}
\centering
\subfloat[\label{fig:Cr16}]{%
       \includegraphics[width=0.25\textwidth]{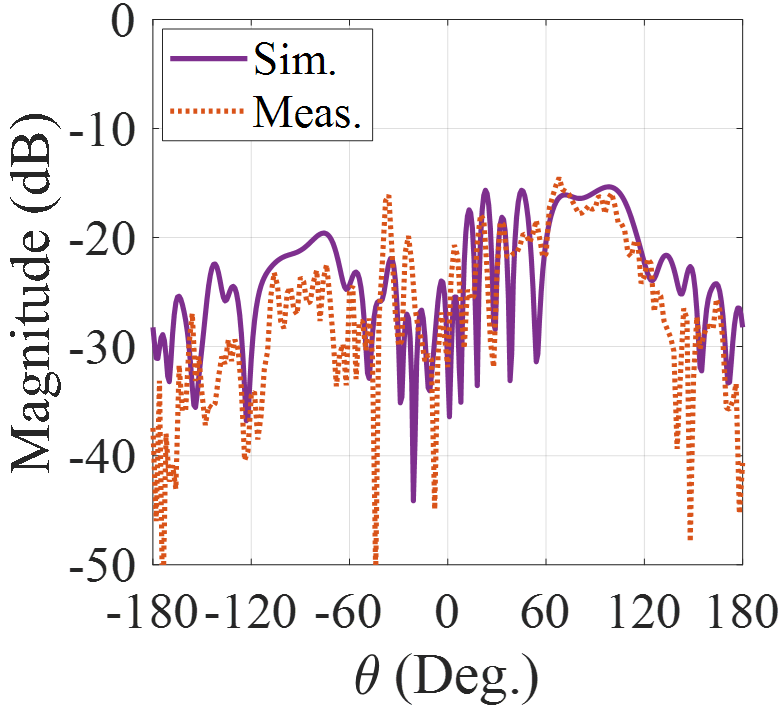}}
%\subfloat[\label{fig:Cr17}]{%
%       \includegraphics[width=0.25\textwidth]{Cr17.png}}
\subfloat[\label{fig:Cr18}]{%
       \includegraphics[width=0.25\textwidth]{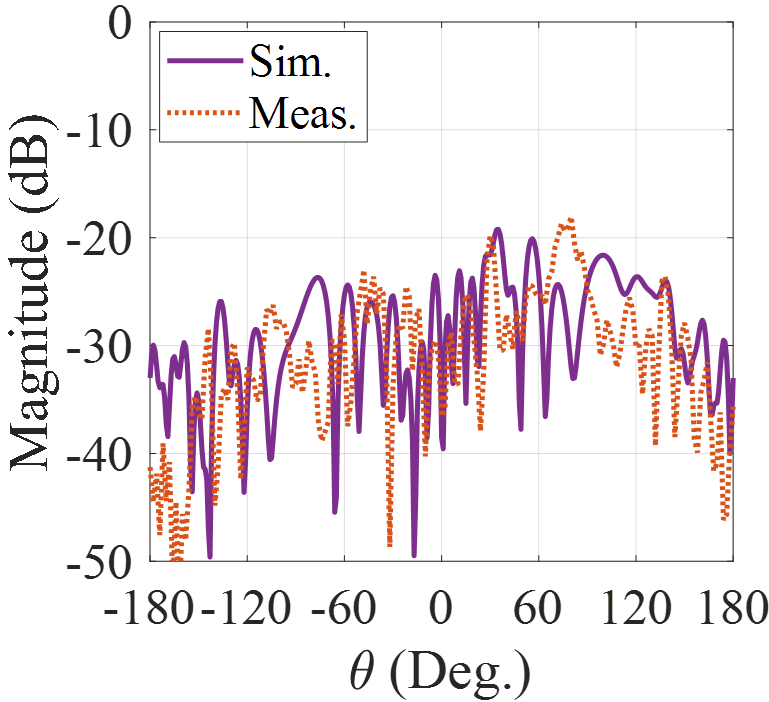}}
\subfloat[\label{fig:Cr19}]{%
       \includegraphics[width=0.25\textwidth]{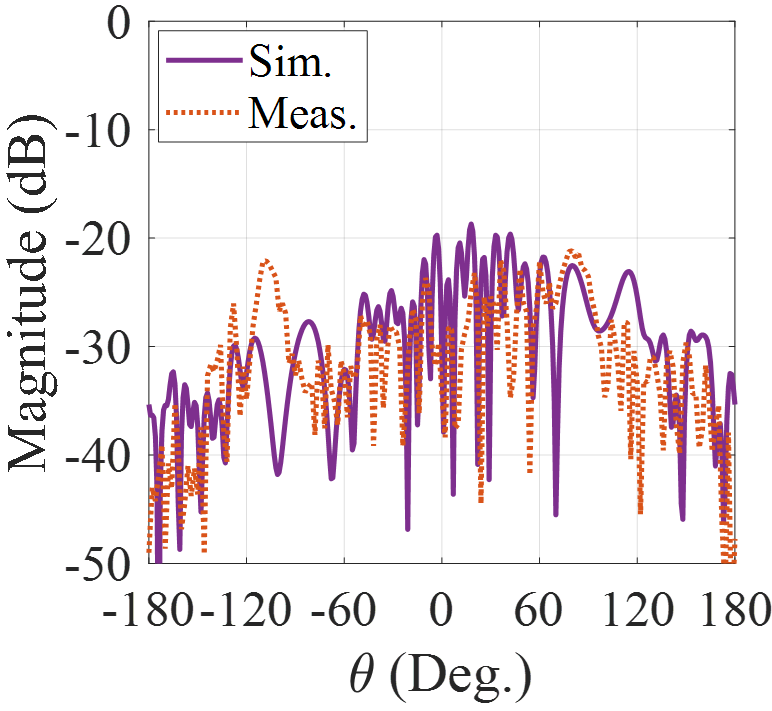}}
%\subfloat[\label{fig:Cr20}]{%
 %      \includegraphics[width=0.25\textwidth]{Cr20.png}}
\subfloat[\label{fig:Cr21}]{%
       \includegraphics[width=0.25\textwidth]{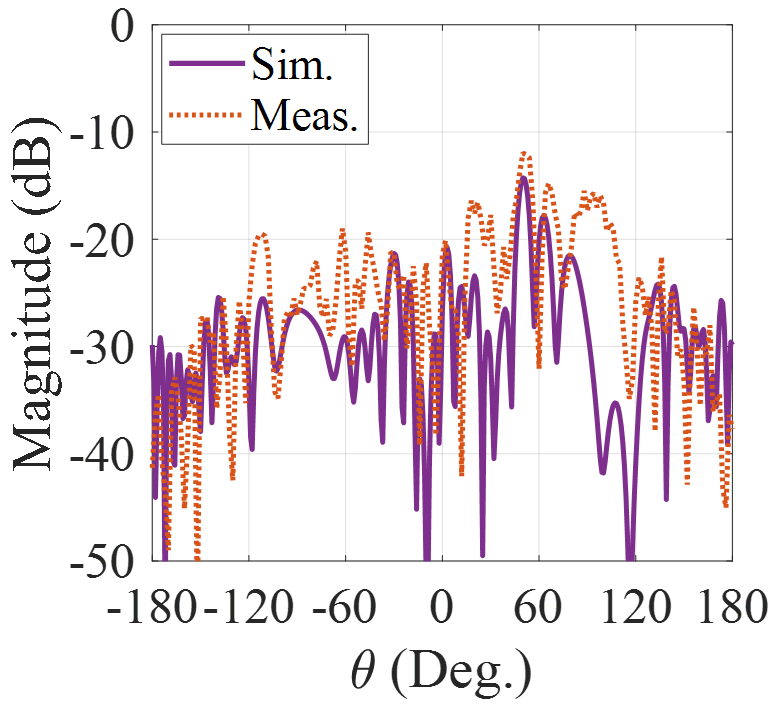}}
\caption{Comparison between simulation and measurement results of RHCP component of far-field patterns at (a) 16 GHz, (b) 18 GHz, (c) 19 GHz and (d) 21 GHz.}
\label{fig:Cr_MEAS}
\end{figure*}

Fig.\ref{fig:Antenna_Exploded} shows the overall structure of antenna and polarizer. The dimensions of the antenna are equal to $145\times105mm^2$ ($\approx 9\lambda \times 6\lambda$). Considering the available laboratory facilities and to reduce the fabrication costs, the dimensions of antenna are selected as small as possible. 
Since the polarizer is designed with assumption of exciting by ($\vec{E}^{inc}$) in (\ref{eq:Einc}), it is required to put the polarizer on the antenna  in a way that the linear radiated wave from the antenna and the $\vec{E}^{inc}$ be exactly at the same direction. 
Ring-shaped spacers are implemented at the corners and edges of structure to create the required air gaps. Since most of the radiated power is focused on the central region of structure, these spacers do not affect the overall performance of antenna.
In Fig.\ref{fig:Pattern_Scan_Circ}, the simulation results of 3-D patterns for co- and cross-polarization components at different frequencies are illustrated. They show that the gain changes from 18 to 20.4 dB in this bandwidth. Also, the side lobe level in scan plane is better than -11 dB across the bandwidth. The details of far-field parameters are given in Table \ref{tab:tab3}.
\begin{table}[]
\caption{Simulation results of proposed antenna in Fig.\ref{fig:Antenna_Exploded}}
\label{tab:tab3}
\begin{tabular}{c|ccccc}
\hline\hline
\begin{tabular}[c]{@{}c@{}}Freq.\\ (GHz)\end{tabular} & \begin{tabular}[c]{@{}c@{}}Beam Direction\\ (Deg.)\end{tabular} & \begin{tabular}[c]{@{}c@{}}LHCP Gain\\ (dB)\end{tabular} & \begin{tabular}[c]{@{}c@{}}RHCP Gain\\ (dB)\end{tabular} & \begin{tabular}[c]{@{}c@{}}SLL\\ (dB)\end{tabular} & \begin{tabular}[c]{@{}c@{}}AR\\ (dB)\end{tabular} \\ \hline
16                                                    & 19                                                              & 18                                                       & 2.7                                                      & -11.1                                              & 1.10                                              \\
17                                                    & 25                                                              & 18.8                                                     & 1.8                                                      & -13.4                                              & 1.16                                              \\
18                                                    & 30                                                              & 19.9                                                     & 0.7                                                      & -14.1                                              & 1.45                                              \\
19                                                    & 36                                                              & 20.4                                                     & 1.7                                                      & -13.3                                              & 1.12                                              \\
20                                                    & 42                                                              & 20.3                                                     & -1.1                                                     & -13.9                                              & 0.69                                              \\
21                                                    & 47                                                              & 19.6                                                     & 5.3                                                      & -11.0                                              & 2.41                                              \\ \hline
\end{tabular}
\end{table}
To verify the simulation results, a prototype of antenna is fabricated and measured in an anechoic chamber. The patches and the metal layers used on the dielectric substrate are made of copper. Also, metalized holes with diameter of 0.5 mm are drilled to realize the metallic vias.
In Fig.\ref{fig:ُProt_S11} the fabricated antenna, the measurement setup and its measured voltage standing ratio (VSWR) are depicted. The impedance bandwidth of antenna is obtained between 16 to 21 GHz. In Figs \ref{fig:Co_MEAS} and \ref{fig:Cr_MEAS} the simulation and measurement results of far-field patterns at different frequencies for both co- and cross-polarizations are compared.

\begin{figure*}
\centering
\subfloat[\label{fig:AR16}]{%
       \includegraphics[width=0.25\textwidth]{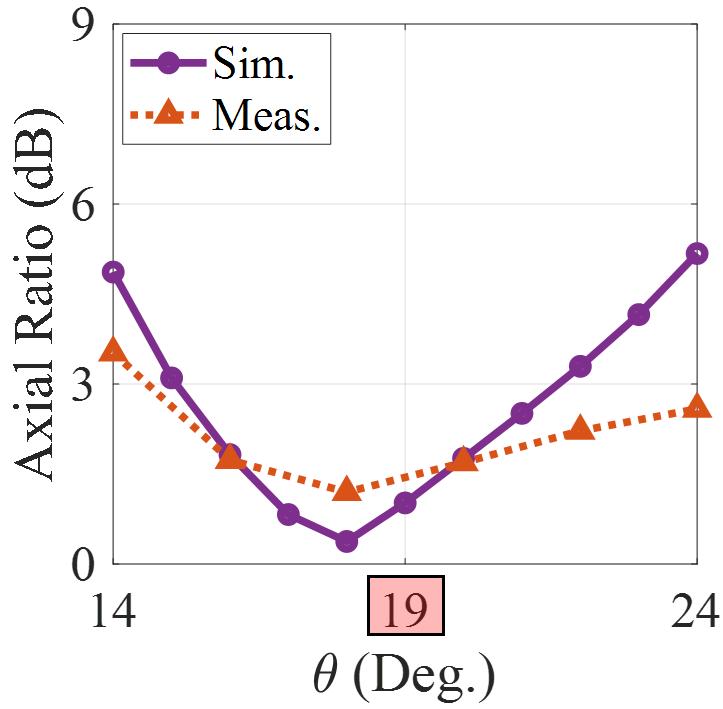}}
%\subfloat[\label{fig:Cr17}]{%
%       \includegraphics[width=0.25\textwidth]{Cr17.png}}
\subfloat[\label{fig:AR18}]{%
       \includegraphics[width=0.25\textwidth]{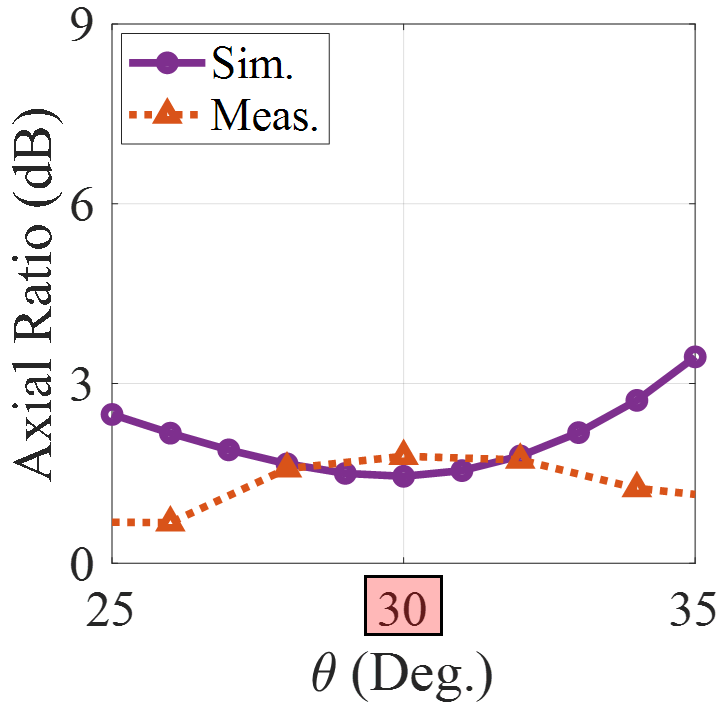}}
\subfloat[\label{fig:AR19}]{%
       \includegraphics[width=0.25\textwidth]{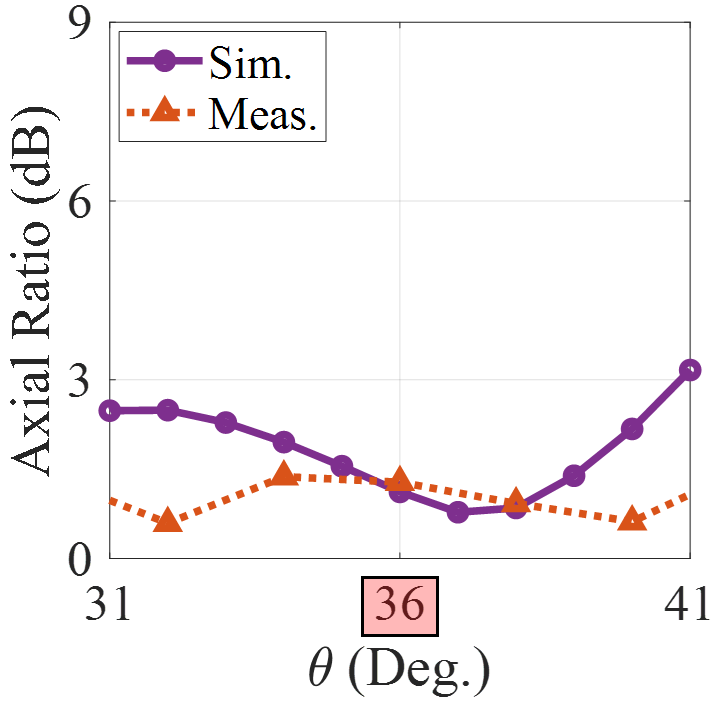}}
%\subfloat[\label{fig:Cr20}]{%
 %      \includegraphics[width=0.25\textwidth]{Cr20.png}}
\subfloat[\label{fig:AR21}]{%
       \includegraphics[width=0.25\textwidth]{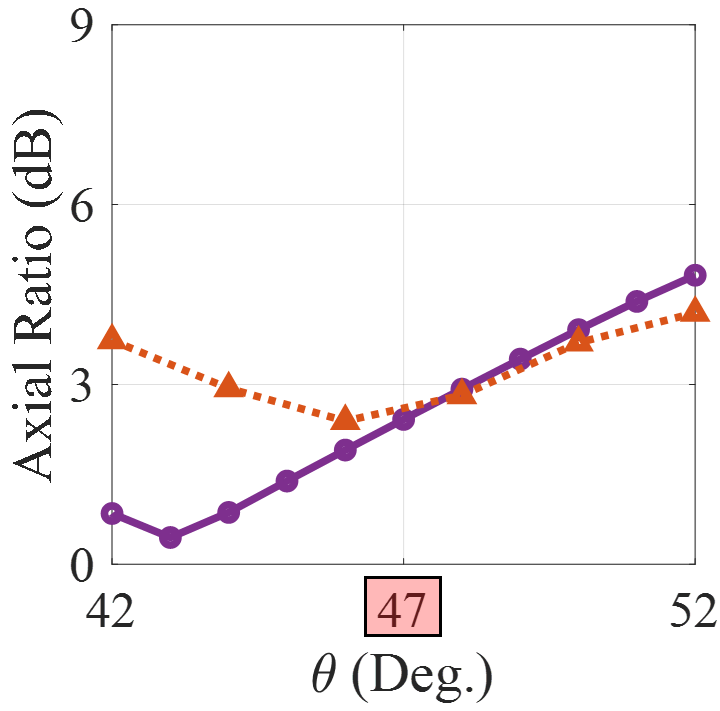}}
\caption{Comparison between simulation and measurement results of axial ratio in the far-field at (a) 16 GHz, (b) 18 GHz, (c) 19 GHz and (d) 21 GHz.}
\label{fig:AR_MEAS}
\end{figure*}
 \begin{table*}
 \centering
\caption{Summery of recent development of circularly polarized MoMetAs and comparison between them.  }
\label{tab:table4}
\begin{tabular}{c|cccccc}
\hline\hline
Ref.      & Synthesis Method & Features                 & Unit-cell Type & \begin{tabular}[c]{@{}c@{}}Frequency BW \\ (GHz)\end{tabular} & \begin{tabular}[c]{@{}c@{}}Co-to-cross Pol. Ratio\\ (dB)\end{tabular} & Scannability \\ \hline
{[}11{]}  & Holography       & Center-fed Hologram      & Anisotropic    & 10                                                            & 19.6                                                                  & No           \\
{[}17{]}  & AFE              & Spiral MoMetA            & Anisotropic    & N.A.                                                          & $\sim$12.5                                                            & No           \\
{[}17{]}  & AFE              & Multi-beam Spiral MoMetA & Anisotropic    & 19.5-20.5                                                     & $\sim$17-20                                                           & No           \\
{[}18{]}  & AFE+MoM          & Spiral MoMetA            & Isotropic      & 17                                                            & 10                                                                    & No           \\
{[}18{]}  & AFE+MoM          & Spiral MoMetA            & Anisotropic    & 8.4521                                                        & $\sim$25                                                              & No           \\
{[}19{]}  & Flat Optics      & Spiral MoMetA            & Anisotropic    & 26.25                                                         & N.A.                                                                  & No           \\
{[}34{]}  & Holography       & Center-fed Hologram      & Anisotropic    & 10                                                            & $\sim$6                                                               & No           \\
{[}34{]}  & Holography       & Dual-beam Center-fed     & Anisotropic    & 9.5-11                                                        & $\sim$7-9                                                             & Yes          \\
This Work & AFE              & Reflector-based MoMetA   & Anisotropic    & 16-21                                                         & 14.3-21.4                                                             & Yes          \\ \hline
\end{tabular}
\end{table*}
 The results show that the simulations and the measurments are in a good agreement. In Fig.\ref{fig:AR_MEAS}, the simulation and measured results of axial ratio parameters are illustrated. The axial ratio at each frequency is measured at an angle which has the maximum radiation gain. Observe that axial ratios at angles with the maximum radiation are less than 3 dB, indicating an acceptable circularly polarized radiation.
 In Table \ref{tab:table4}, the characteristics of proposed antenna are compared with the circularly polarized MoMetAs in other references. Observe that the cross-polarization level of the proposed isotropic antenna in reference \cite{minatti2016} is not suitable, and thus anisotropic structure is used instead to resolve this issue. Moreover, except for reference \cite{wan2016_AOM} and our proposed antenna, all other structures are not capable of frequency scanning. Although the proposed antenna in \cite{wan2016_AOM} has scannability, its cross-polarization level is not desired. In \cite{teniou2017}, also the antenna has a bandwidth spectrum of 19.5 to 20.5 GHz, yet it has a constant radiation beam across its bandwidth. 
\section{Conclusion}
In this paper, a surface-wave reflector is used to incorporate the scannability to an anisotropic holographic antenna with vertical polarization. The design method is based on the combination of holography technique and aperture field estimation theory. The merits of this method are its capability to synthesize different polarizations and to estimate the radiation pattern with acceptable accuracy, without the need for full-wave simulations, which accelerates the synthesis and design procedure. To realize the holographic antenna, asymmetric rectangular-shaped unit-cells with size of $\lambda/5.1$ with respect to the wavelength at the upper limit of design frequency (21 GHz) are used, which effectively model the surface impedance tensor. To achieve an acceptable polarization purity, both the size of patches and their rotation angle are utilized as the design parameters. The fractional bandwidth of antenna with circular polarization is obtained up to 27\%. The circularly polarized radiation is enabled by a wide-band three-layer polarizer with high angular stability. The designed polarizer does not deteriorate the performance of antenna. Subsequently, the final structure has the same bandwidth of holographic antenna, while it has circular polarization. The radiation pattern of the antenna is approximately pencil-beam, making it suitable for telecommunication systems and radars with high resolutions.


%\newpage


%\vfill


\end{document}